\documentclass[11pt,preprint]{aastex6}

\usepackage{graphics}
\usepackage{epsfig}
\usepackage{graphicx}
\usepackage{color}
\usepackage{pstricks}
\usepackage{afterpage}
\usepackage{epstopdf}
\usepackage{natbib}

\newcommand{\unit}[1]{\ifmmode\,{\rm #1}\else$\,{\rm #1}$\fi}

\newcommand{\etal}{~{et~al.}\ }  

\def\kms{km~s$^{-1}$}
\def\arcmin{$^{\prime}$}
\def\arcsec{$^{\prime\prime}$}

\newcommand{\sqd}{deg$^{2}$}
\newcommand{\hi}{H{\sc\,i}}

\usepackage{natbib}

\usepackage{longtable}

\slugcomment{Draft \today}

\shorttitle{ALFALFA Extragalactic \hi\ Catalog}
\shortauthors{Haynes\etal}

\begin{document}

\title{The Arecibo Legacy Fast ALFA Survey: The ALFALFA Extragalactic HI Source Catalog}
\author{
Martha P. Haynes\altaffilmark{1},
Riccardo Giovanelli\altaffilmark{1},
Brian R. Kent\altaffilmark{2},
Elizabeth A.K. Adams\altaffilmark{3,4},
Thomas J. Balonek\altaffilmark{5},
David W. Craig\altaffilmark{6},
Derek Fertig\altaffilmark{7},
Rose Finn\altaffilmark{8},
Carlo Giovanardi\altaffilmark{9},
Gregory Hallenbeck\altaffilmark{10},
Kelley M. Hess\altaffilmark{4,3},
G. Lyle Hoffman\altaffilmark{11},
Shan Huang\altaffilmark{1},
Michael G. Jones\altaffilmark{12},
Rebecca A. Koopmann\altaffilmark{14},
David A. Kornreich\altaffilmark{1},
Lukas Leisman\altaffilmark{13},
Jeffrey Miller\altaffilmark{15},
Crystal Moorman\altaffilmark{16},
Jessica O'Connor\altaffilmark{17},
Aileen O'Donoghue\altaffilmark{15}, 
Emmanouil Papastergis\altaffilmark{4,18},
Parker Troischt\altaffilmark{19},
David Stark\altaffilmark{20}
Li Xiao\altaffilmark{21}
}
\altaffiltext{1}{Cornell Center for Astrophysics and Planetary Science, 
Space Sciences Building, Cornell University, Ithaca, NY 14853, USA; haynes@astro.cornell.edu}
\altaffiltext{2}{National Radio Astronomy Observatory, 520 Edgemont Rd., Charlottesville,
  VA 22901, USA}
\altaffiltext{3}{ASTRON, Netherlands Institute for Radio Astronomy, Postbus 2, 7900 AA, Dwingeloo, The Netherlands}
\altaffiltext{4}{Kapteyn Astronomical Institute, University of Groningen, 
  Landleven 12, 9747 AD, Groningen, The Netherlands}
\altaffiltext{5}{Department of Physics and Astronomy, Colgate University, Hamilton, NY 13346, USA}
\altaffiltext{6}{West Texas A\&M University Department of Chemistry and Physics, 2403 Russell Long Blvd.
  Canyon, TX 79015, USA}
\altaffiltext{7}{Langley High School, 6520 Georgetown Pike, McLean, VA 22101, USA}
\altaffiltext{8}{Physics Department, Siena College, Loudonville, NY 12211, USA}
\altaffiltext{9}{Osservatorio Astrofisico di Arcetri, Largo E. Fermi 5, I-50125 Firenze, Italy}
\altaffiltext{10}{Washington and Jefferson College, Department of Computing and Information Studies, 60 S Lincoln Street, Washington PA, 15301, USA}
\altaffiltext{11}{Department of Physics, Hugel Science Center, Lafayette College, Easton, PA  18042, USA}
\altaffiltext{12}{Instituto de Astrof\'isica de Andaluc\'ia, CSIC, Glorieta de la Astronom\'ia s/n
  E-18008, Granada, Spain}
\altaffiltext{13}{Department of Physics and Astronomy, Valparaiso University, Neils Science Center, 1610 Campus Drive East,
Valparaiso, IN 46383, USA}
\altaffiltext{14}{Department of Physics and Astronomy, Union College, 807 Union Street,
  Schenectady, NY 12308, USA}
\altaffiltext{15}{St. Lawrence University, 23 Romoda Drive, Canton, NY 13617, USA}
\altaffiltext{16}{Department of Phyics, Lynchburg College, 1501 Lakeside Drive, Lynchburg, VA 24501, USA}
\altaffiltext{17}{Department of Physics and Astronomy, George Mason University, 4400 University Drive, MSN: 3F3, Fairfax, VA 22030, USA}
\altaffiltext{18}{Credit Risk Modeling Department, Coöperative Rabobank U.A., Croeselaan 18, Utrecht NL-3521CB, The Netherlands}
\altaffiltext{19}{Department of Physics, Hartwick College, Oneonta, NY 13820, USA}
\altaffiltext{20}{Kavli Institute for the Physics and Mathematics of the Universe (WPI), The University 
of Tokyo Institutes for Advanced Study, The University of Tokyo, Kashiwa, Chiba 277-8583, Japan}
\altaffiltext{21}{National Astronomical Observatories, Chinese Academy of Sciences, 20A Datun Road, Chaoyang District, Beijing 100012, China}
\begin{abstract}
We present the catalog of $\sim$31500 extragalactic HI line sources detected by
the completed ALFALFA survey out to z $<$ 0.06 including both high signal-to-noise
ratio ($>$ 6.5) detections and ones of lower quality which coincide in both position and recessional
velocity with galaxies of known redshift. We review the observing technique, 
data reduction pipeline, and catalog construction process, focusing on details of particular 
relevance to understanding the catalog's compiled parameters. We further describe and 
make available the digital HI line spectra associated with the catalogued sources. 
In addition to the extragalactic HI line detections, 
we report nine confirmed OH megamasers and ten OH megamaser candidates 
at 0.16 $< z <$ 0.22 whose 
OH line signals are redshifted into the ALFALFA frequency band. 
Because of complexities in data collection
and processing associated with the use of a feed-horn array on a complex
single-dish antenna in the terrestrial radio frequency interference environment, 
we also present a list of suggestions and caveats for consideration by users 
of the ALFALFA extragalactic catalog for future scientific investigations.
\end{abstract}

\section{Introduction}
\label{sec:intro}

\hi\ 21 cm line surveys provide a census of the extragalactic population of atomic gas-bearing galaxies. Because
of the relatively simple physics involved in most \hi\ line emission, conversion of the observed line flux into 
atomic hydrogen gas mass is straightforward, and the spectral nature of the emission provides observable measures
of the redshift and projected disk rotational velocity. While the molecular H$_2$ gas tends to concentrate
in a small number of giant gas clouds principally in the inner regions, the HI disk traces the full extent of
the gas layer. Star formation is linked more closely to
the molecular H$_2$ gas \citep[e.g.][]{kennicutt12a, saintonge16a, catinella18a}. However, in most galaxies, the
HI fills a much larger fraction of 
interstellar space and contributes most of the cool gas mass, thus
representing the fuel reservoir and potential for future star formation. 

The Arecibo Legacy Fast ALFA (ALFALFA) Survey used the seven-horn Arecibo L-band Feed Array (ALFA)
to map nearly 7000 \sqd~ of high Galactic latitude sky accessible to the Arecibo telescope over $\sim$4400 
nighttime hours between 2005 and 2011. ALFALFA was conducted as a ``blind'' survey: at each position, the entire  
frequency range from 1335-1435 MHz, corresponding to heliocentric velocities -2000 $<$ c$z <$ 18000 \kms,
was searched for line emission. 
As described in detail by \citet{giovanelli05a}, the 
ALFALFA survey design was largely dictated by the principal science goal of determining the
faint end of the HI mass function (HIMF),
and the overall abundance of low mass gas-rich halos
\citep[e.g.][]{martin10a, papastergis11a, martin12a, papastergis13a}. Additional
objectives include: how the HIMF might vary with environment 
\citep[e.g.][]{moorman14a, jones16b, jones18b}, how the HI-bearing population differs
from optically-selected ones \citep[e.g.][]{huang12b, huang12a, gavazzi13a},
using the HI distribution to look for 
tidal debris on large angular scales \citep[e.g.][]{lee-waddell14a, lee-waddell16a, leisman16a},
and establishing metrics for the normal HI content of galaxies
\citep[e.g.][]{toribio11a, odekon16a}. As the least clustered local ($z \sim$ 0) galaxy population
\citep{martin12a}, the HI-bearing population traces how galaxies evolve when left
on their own, in relative isolation.

ALFALFA has also discovered an number of enigmatic objects such as the nearby 
faint dwarf Leo P \citep{giovanelli13a}, the very metal poor Leoncino 
\citep{hirschauer16a} and the highly HI-dominated Coma P \citep{janowiecki15a, ball18a}.
Additionally, ALFALFA has provided the opportunity to survey classes of galaxies 
such as extremely low HI mass dwarfs \citep[e.g.][]{cannon11a, teich16a, mcnichols16a}, 
galaxies with extremely high HI-to-stellar mass ratios  \citep[e.g.][]{adams15b, janowiecki15a,
janesh15a, janesh17a} and 
HI-bearing ultra diffuse galaxies \citep{leisman17a}. The vast majority ($>98$\%) of 
extragalactic ALFALFA sources can be associated with at least one likely
stellar counterpart, and the majority of the ``dark'' objects are
likely associated with tidal debris in interacting systems \citep[e.g.][]{haynes07a, 
koopmann08a, lee-waddell14a, leisman16a}. A few dark galaxy candidates remain 
intriguing, and continuing work seeks to identify associated starlight and
constrain their dynamics and star formation history \citep[e.g.][]
{kent10a, giovanelli10a, cannon15a}.

As a complement to \citet{jones18b} which presents the derived HIMF and its dependence
on local environment,
this paper presents the extragalactic HI catalog extracted from the completed ALFALFA survey.
Because of the overlapping nature of the drift scan survey and improved availability
of the public optical imaging used to identify optical counterparts (OCs) of ALFALFA
HI sources, this catalog, presented in Table \ref{tab:exgalcat}, supersedes and
replaces previous releases \citep{giovanelli07a, saintonge08a, kent08a, stierwalt09a,
martin09a, haynes11a}. In addition to the catalog of ALFALFA HI line detections,
nine sources are identified with OH megamasers (OHMs) and ten are flagged as being
OHM candidates. 

Section \ref{sec:survey} reviews the important aspects of the ALFALFA survey
observational program and data reduction process which has led to the
production of the extragalactic dataset presented in Section \ref{sec:catalog}. Section
\ref{sec:summ} summarizes a number of important points, realities and caveats
about the survey and its
resultant data products which readers are encouraged to keep in mind. Appendix
\ref{app:pipeline} presents details of the data acquisition and processing pipeline
used to produce the ALFALFA catalog.

To allow direct comparison with the vast majority of extant works on HI line
emission at low redshift, we use the observed rest frame and do not apply cosmological
corrections dependent on redshift; those amount to at most a few percent for the most
distant sources. Details of this choice are given in the text.

\section{The ALFALFA Survey}
\label{sec:survey}

\begin{figure}[t!]
\centering
\includegraphics[height=2.1in]{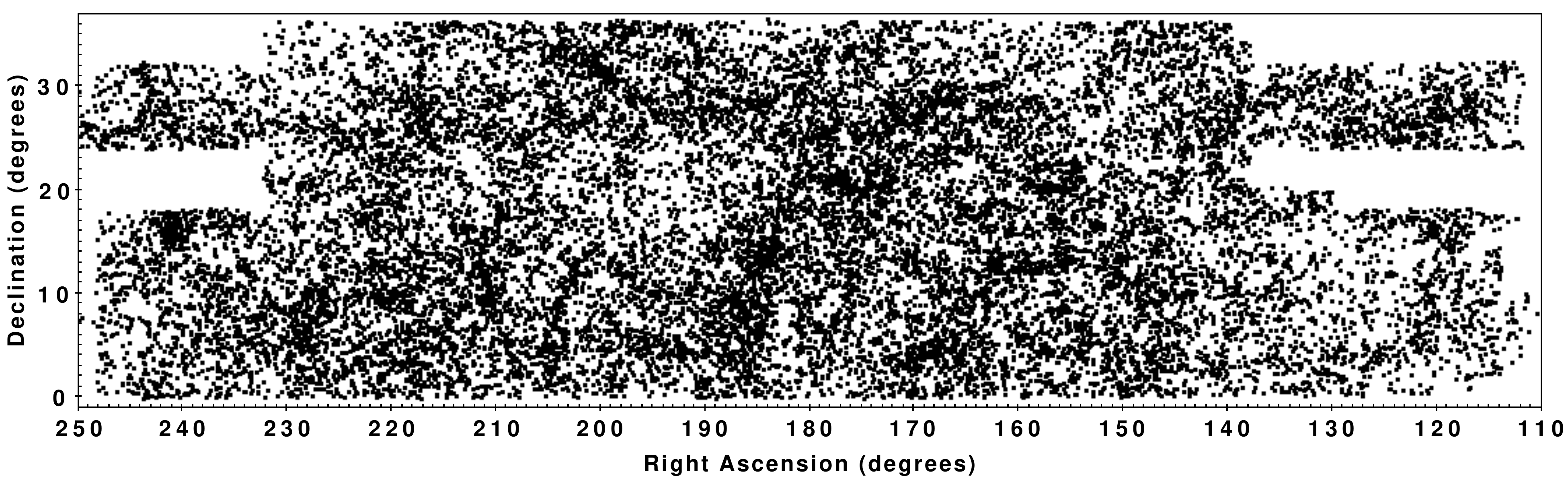}\\
\includegraphics[height=2.1in]{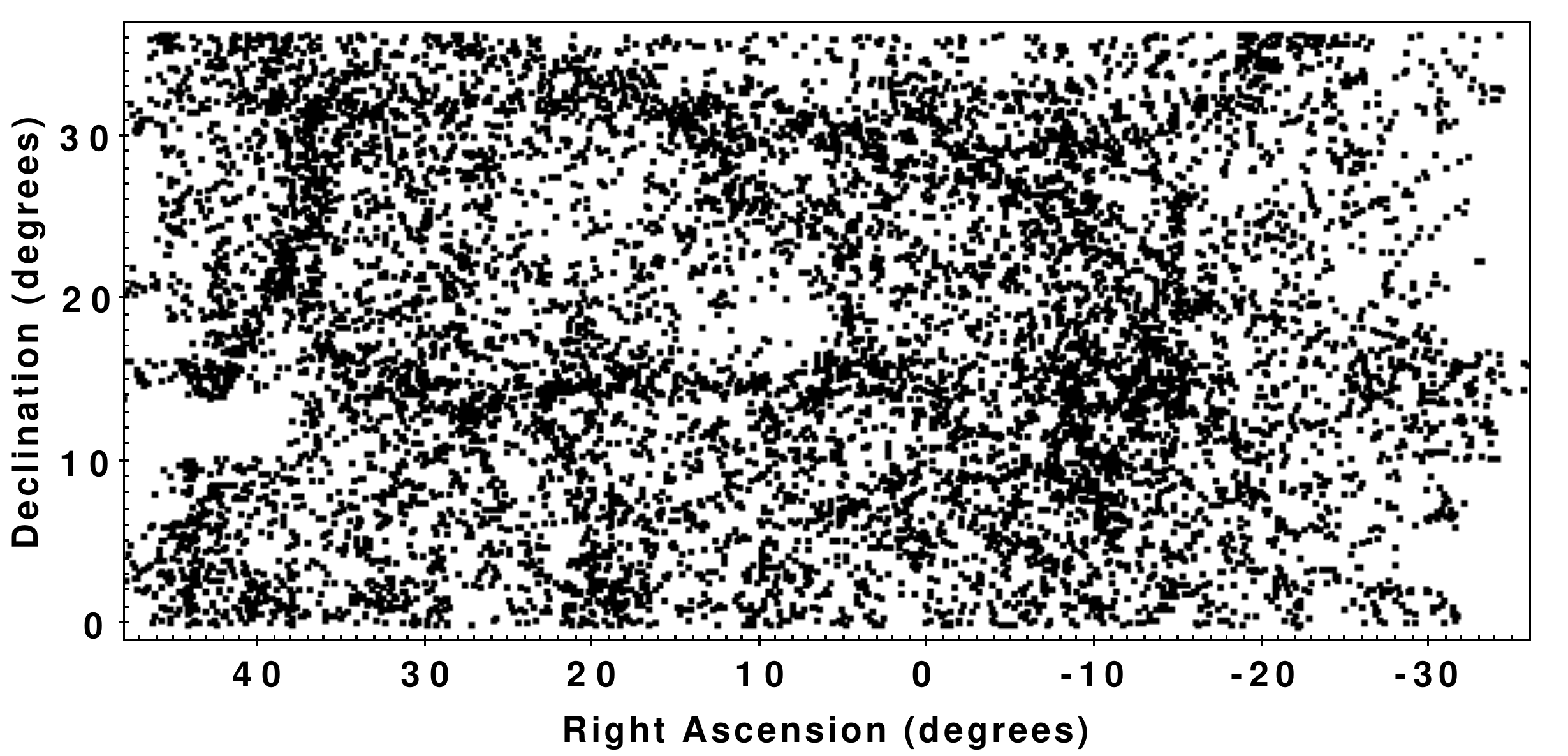}
\vspace{-0.1cm}
\caption{Sky distribution of ALFALFA sources included in Table \ref{tab:exgalcat} in
the northern (top) and southern Galactic hemispheres (bottom), showing the roughness
of boundaries imposed by practical and scheduling constraints.}
\label{fig:skyplot}
\end{figure}

The ALFALFA survey was intended to cover two sky areas at high Galactic latitude, one
in the northern Galactic hemisphere
07$^h$30$^m$ $<$ R.A. $<$ 16$^h$30$^m$,  
0$^{\circ} < $ Dec. $<$ +36$^{\circ}$ 
and one in the southern hemisphere, 
22$^{h}$ $<$ R.A. $<$ 03$^{h}$, 0$^{\circ} <$ Dec.
$<$ +36$^{\circ}$. For various practical reasons, the final sky area, depicted here in
Figure \ref{fig:skyplot}, also shown in Figure 1 of
\citet{jones18b}, is somewhat reduced near the edges. 

As discussed in \citet{giovanelli05a} and \citet{giovanelli15a}, the ALFALFA survey was designed particularly to
sample the HIMF over a fair cosmological distance of $\simeq$100 Mpc, therefore setting minimum requirements on the
survey volume sensitivity and areal coverage. For a survey with a telescope
characterized by a given system temperature T$_{sys}$ and gain G, the science-driven need to detect
a given HI mass $M_{HI}$ of HI line width $W_{50}$ \kms~ at a distance $D_{Mpc}$ translates to a required
integration time $t_{int}$ in seconds
\begin{equation}
t_{int} \propto (T_{sys}/G)^2 ~M_{HI}^{-2} ~D_{Mpc}^4 ~W_{50}^{-2\gamma}
\end{equation}
where $\gamma  \simeq -1/2$ for $W_{50} < 200$ \kms,
 $\gamma \simeq -1$ for $W_{50} > 200$ \kms \citep{giovanelli05a, giovanelli15a}.
The ALFALFA HIMF science goal dictates that the survey cover a very wide solid angle $\Omega_{survey}$ $\sim$ 7000 \sqd\
with an average integration time of $\sim$48 seconds per beam solid angle after combination of all drifts from all beams
and polarizations across each spatial point. The sheer amount of telescope time (thousands of hours)
needed to accomplish such wide sky coverage in turn demanded an observing strategy that exploited
Arecibo's large collecting area, the mapping capability of the ALFA instrument and the spectral power of
its backend spectrometer to maximum observing efficiency. 

\subsection{Drift scan technique}\label{sec:driftscan}

As discussed in detail in \citet{giovanelli05a}, ALFALFA was conducted as a drift scan survey using the 7-feed horn array ALFA.
The ALFA feed horn configuration delivers a central, higher gain beam surrounded by a ring of 6 equally-spaced somewhat lower gain beams. 
For most of the survey, the azimuth arm of the telescope was positioned on the meridian at a pre-assigned J2000.0 declination, 
with a spacing of 14.6\arcmin\ between primary drift centers.
The feed array  was rotated by 19 degrees so that the Earth-rotation drift-scan tracks of 
individual beams were equally spaced by 2.1\arcmin\ in declination in J2000.0 coordinates. 
A second, parallel drift pass of the same region of the
sky was acquired later, with the center beam offset from the first by 7.3\arcmin ~(half the spacing to the next primary center beam positioning) so that the final sampling in declination was 1.05\arcmin.
Because hardware limits do not allow 
pointing straight overhead, coverage of declinations located close to the zenith (Dec. = +18$^\circ$
21\arcmin) with similar parallel tracks required the telescope to be positioned off-meridian and the array rotated by a
different amount, depending on the declination of the array center.
The spectra were acquired covering a 100 MHz bandwidth centered at 1385 MHz using the WAPP (Wide-band Arecibo Pulsar Processer)
spectrometer, yielding 4096 ``channels'' per spectrum, equally spaced in frequency, for each of two linear polarizations of 
each of the seven feed horns (a total of 14 spectra). 
Parameters of the ALFALFA observing setup and specifications are summarized in Table \ref{tab:observingparms}.

\begin{table*}[]
\caption{\textbf{ALFALFA Technical Details}}
\vskip 5pt
\centering
\footnotesize{
\begin{tabular}{ll}
\hline
\hline
Number of beams                 & 7 \\
Polarizations per beam          & 2 linear \\
Beam size (FWHM)                & 3\arcmin.8~$\times$~3\arcmin.3 \\
Gain                            & 11 K/Jy (central beam) and 8.5 K/Jy (peripheral) \\
$T_{\rm{sys}}$                  & 26-30 Kelvin \\
Frequency range                 & 1335-1435 MHz \\
c$z_{\odot}$ range              & -2000 to 17912 \kms \\
Bandwidth (total)               & 100 MHz \\
Correlator lags (spectral channels)               & 4096 \\
Channel spacing                 & 24.4 kHz (5.1 \kms~ at 1420.4058 MHz) \\
Spectral resolution             & 10 \kms, after Hanning smoothing \\
Autocorrelation sampling               & 3 level \\
Avg. channel rms                & 2.0 mJy/channel \\
Map rms                         & 1.86 mJy/beam \\
Effective map $t_{int}$         & 48 sec (beam solid angle)$^{-1}$\\
5$\sigma$ survey sensitivity    & 0.72 Jy \kms~for W$_{50}$=200 \kms~at $t_{int}$\\
Single drift sky coverage       & 600 sec~$\times$~14.6\arcmin~(all beams) \\ 
Drift scan size on disk         & 213 MB \\
Grid sky coverage               & 2.4~$\times$~2.4 degrees \\ 
Grid center spacing             & 8 min in R.A. and 2$^\circ$ in Dec.\\
Grid c$z_{\odot}$ coverage      & a) $\sim$-2000 to  3300 \kms \\ 
                                & b) $\sim$ 2500 to  7950 \kms \\ 
                                & c) $\sim$ 7200 to 12800 \kms \\  
                                & d) $\sim$12100 to 17912 \kms \\ 
Grid c$z$ overlap               & 140 channels \\ 
Grid size on disk               & 381 MB \\
\hline
\hline
\end{tabular}\label{tab:observingparms}
}
\end{table*}

The drift scan observations 
were conducted in observing runs that typically lasted four to nine hours at 
a time, normally without interruption, yielding an exceptionally high efficiency of ``open-shutter time''. 
Once data acquisition for an observing run began, 14 individual spectra (polarizations/beams) were recorded each second
at 99\% time efficiency except for two adjustments made every 600 seconds. First, minor pointing corrections were
made to maintain the pointing of the ALFA central beam in constant declination J2000.0 coordinates; 
it may be noted that this approach insured that adjacent or contiguous drift scans taken several years apart would 
thus remain parallel in that coordinate frame. The corrections from current epoch to J2000 coordinates 
depend on source position and over the seven-year period of data-taking amounted in some 
positions to several arcminutes.

In addition to the minor position update, the data acquisition
sequence was interrupted every 600 seconds
to allow the injection of a calibration noise diode for one second; because of hardware notifications
(``hand-shaking''), this procedure, described in more detail in Appendix \ref{app:pipeline}, took in practice between 
four and seven seconds, still less than the time for a source to cross a single ALFA beam (14 seconds). 
No other adjustments were made. This ``minimum intrusion'' 
approach allowed tracking of separate polarization/beam/spectral behavior over the timescale of hours to
compensate for systematic variations (e.g., drifts in ``electronic gain''). Occasionally, hardware failures led the observing 
sequence to be aborted. In such cases, power levels were readjusted before data acquisition was restarted. 
Because of the desire to calibrate using a significant number (at least nine) of calibration samples, drift sequences 
of less than 90 minutes
were discarded. In general, the lack of power readjustment and minimal telescope motion delivered very high 
overall data quality and robust system calibration. 

\subsection{Radio frequency interference}\label{sec:rfi}

A major complication of observing the HI 21 cm line in the 1335-1435 MHz range is introduced by the presence of
human-generated radio frequency interference (RFI), typically over relatively narrow ranges in frequency occupying a few MHz or
substantially less. Some RFI is predictable, some is (nearly) omnipresent, and some is transient. Most RFI is polarized and some
is very strong, causing a rise in the system temperature (T$_{sys}$) and sometimes introducing spectral standing waves (due to
multiple reflections/scattering within the Arecibo telescope optical path).  RFI mitigation was addressed 
in several different ways. To make possible the identification of RFI by statistical differences in power levels, a second drift across
each part of the sky was undertaken, typically with the second pass centered halfway between adjacent tracks of the first pass
and acquired 3-9 months later than the first. Since doppler tracking was not implemented, the offset in the time of
data acquisition allowed the discrimination of fixed-frequency RFI from cosmic sources.
Each spot on the sky was included in multiple drift scans, and beams/polarizations, such
that a statistical comparison of subsets of data could be checked for inconsistencies
caused by bursts of RFI.

The strongest and most persistent (except for a period of a few months for its replacement) RFI feature arises from the FAA 
radar at the San Juan airport centered near 1350 MHz. The airport radar transmission  is pulsed, 
polarized, azimuth-dependent and not picked up equally by all beams. When it is strong, harmonics generated within the
Arecibo spectral chain may show up at 1380 and (sometimes) 1405 and 1410 MHz. Another common RFI source, evident in shorter
bursts of 60-180 seconds at a time, is associated with the NUclear DETonation (NUDET) detection system aboard the Global 
Positioning System (GPS) satellites. Many other transient RFI sources were present, arising from spurious transmissions,
faulty equipment, etc. 
In order to address RFI contamination, 
each individual polarization/beam spectral drift scan was run through an RFI flagging routine and then examined by an expert who could 
accept or reject the pipelined flags and/or set additional ones. While laborious, this procedure of data flagging produced a 
spectral mask which maintains a record of flagged spectral pixels, important for identifying RFI ``holes'' in the 21 cm line sky,
as the spectrum at each grid point is associated with a spectral weight at each frequency/velocity point.
 
\begin{figure}[t!]
\centering
\includegraphics[height=2.1in]{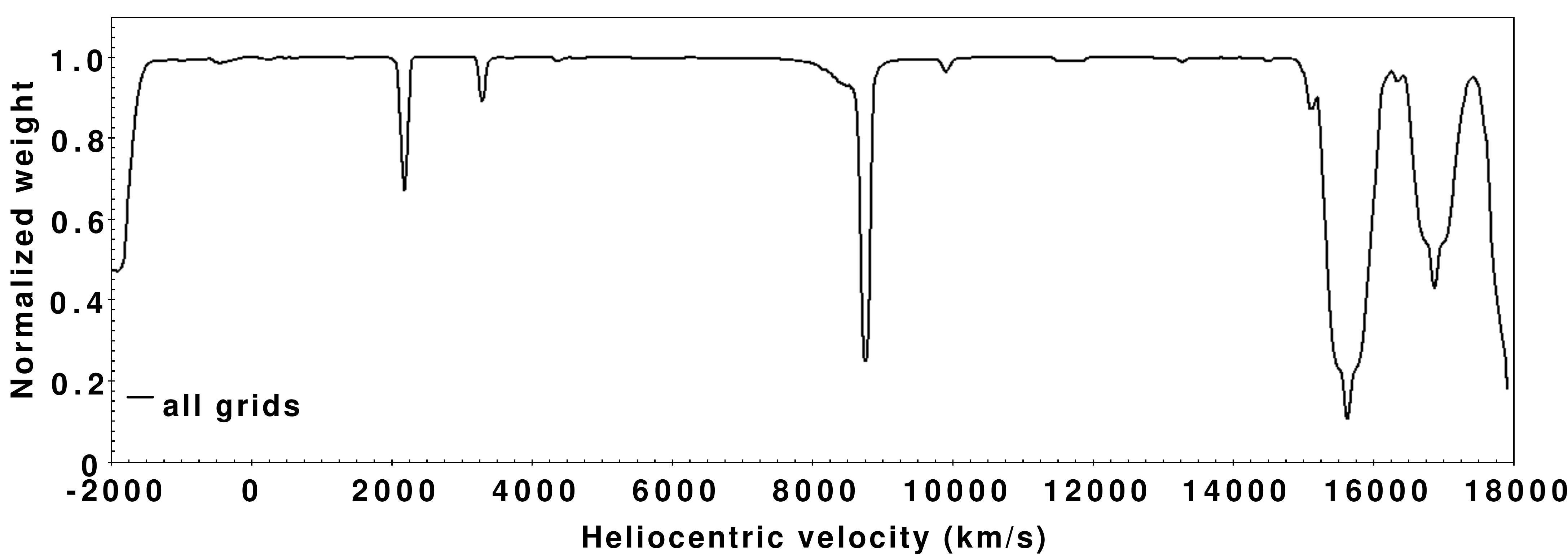}\\
\includegraphics[height=2.1in]{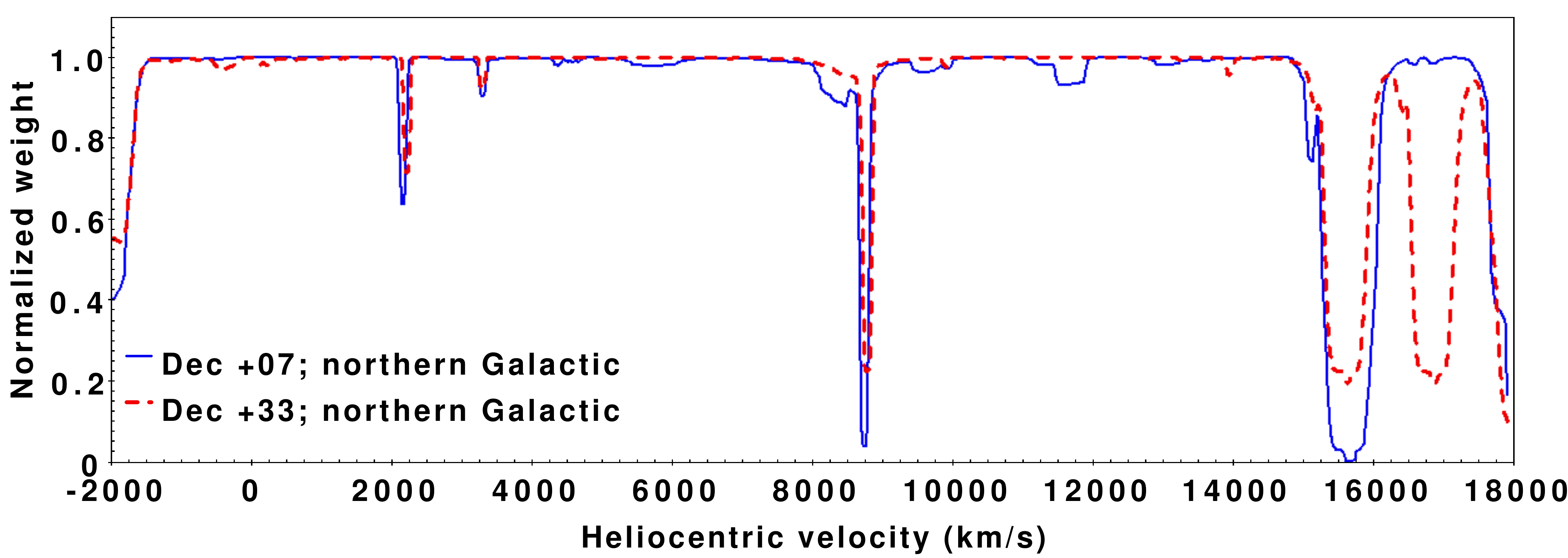}
\vspace{-0.1cm}
\caption{
The normalized spectral weight at each of the final (non-overlapping) 3672 frequency channels 
displayed in the corresponding velocity units for
the entire set of ALFALFA grids (upper) and for two strips of grids across the northern Galactic hemisphere (lower):
at Dec. = +07$^\circ$ (blue line) and +33$^\circ$ (dashed red line). The main cause of missing data (reduced spectral weight) 
is RFI, most notably the 1350 MHz (c$z \sim$15600 km/s) FAA radar at the San Juan airport. Narrower features at 8800, 
3300 and 2200 km/s are modulations of the FAA radar occurring within the WAPP spectrometer. Comparison of the panels illustrates 
the varying nature of the RFI during different observing periods and the serious contamination caused by the 1345-1350 MHz radar
system at velocities above 15000 \kms.
}
\label{fig:gridrfi}
\end{figure}

Similar to the depictions of the typical spectral weights in previous ALFALFA data release papers,
e.g., Figure 1 of \citet{giovanelli07a} and Figure 6 of \citet{martin10a},
Figure \ref{fig:gridrfi} shows the normalized weight per spectral channel derived from the
entire set of ALFALFA grids (top panel) and for two different declination strips of grids covering the northern Galactic
ALFALFA regions (bottom panel). The most prominent reduced-weight features reflect contamination by the San Juan airport 
FAA radar near 1350 MHz ($\sim$15600 \kms)and modulations of it at 1380 MHz ($\sim$8800 \kms), 
1405 MHz ($\sim$3300 \kms), and 1410 MHz ($\sim$2200 \kms).
As found earlier by \citet{giovanelli07a},
on average, about 85\% of the total bandpass was RFI-free with normalized weight $>$ 0.9. 94\% of the bandpass carries
spectral weight of $>$ 0.5; that value can serve as an acceptable limit on data quality. Because of the large percentage
of channels corrupted by the FAA radar systems at frequencies below 1350 MHz,
statistical studies requiring volume completeness 
should be restricted to galaxies within the corresponding velocity limit of c$z <$ 15000 \kms~
\citep{martin10a}.

A more detailed discussion of the drift scan data acquisition, calibration,  processing and RFI flagging process
is presented in Appendix \ref{app:pipeline}.

\subsection{Grid production}\label{sec:gridproduction}

Upon acquisition of all the drift scans covering a region of sky, all of the relevant spectra were combined to
produce a 3-dimensional spectral grid; further details of this process are given in Appendix \ref{app:gridapp}. 
In the spatial domain, standard ALFALFA grids are 
2.4$^{\circ}$ $\times$ 2.4$^{\circ}$, evenly sampled at 
1\arcmin, so that the spatial dimensions of a grid are 144 $\times$ 144. Grid centers are pre-determined, separated
by 8 minutes in R.A. (e.g., 23$^h$00$^m$, 23$^h$08$^m$, 23$^h$16$^m$, etc) and 2$^{\circ}$ in Dec. from +01$^{\circ}$ to +35$^{\circ}$.
In order to keep grid files small enough to be processed and analyzed on typical 2005-era desktops, four separate grids
were produced at each grid center covering four separate but partially overlapping frequency ranges corresponding to 
four velocity ranges: -2000 $<$ c$z$ $<$ 3300 \kms,
2500  $<$ c$z$ $<$ 7950 \kms, 7200  $<$ c$z$ $<$ 12800 \kms, and 12100 $<$ c$z$ $<$ 17900 \kms\ (see Table \ref{tab:observingparms}). 
The gridding procedure also produces, for every grid point, a record of all of the drift scans, beams and polarizations which 
contribute to the intensity for each spectral value. While the time for a source to drift across a single ALFA beam is about
14 seconds, the effective integration time after grid construction is typically $t_{int}\simeq 48$ seconds 
per beam solid angle. It can be less where significant data are missing. In order to track data quality, 
a normalized weight is recorded for each spectral value.

In addition, the gridding procedure changes the spectral intensities from Kelvins in antenna temperature
to mJy in flux density, correcting for zenith angle variations in the gain of the telescope. The flux density scale, 
set initially by measuring the power injected by the noise diode (see Section \ref{sec:driftscan} and Appendix \ref{app:pipeline}),
is corroborated by comparing the ALFALFA flux densities of unresolved continuum sources in contiguous grids along the same declination
strip with the catalogued flux 
densities at 1400 MHz of the same sources as reported by the 
NRAO VLA Sky Survey \citep[NVSS;][]{condon98}; such comparison typically
involves hundreds of sources over a spring strip of grids centered at the same declination.
If the discrepancy in measured fluxes was significant, all fluxes in the involved grids were corrected 
by a small multiplicative factor to bring them in line with the NVSS values. 
In no case was the average continuum offset found to be greater than 4\% and usually 
was within 2\%.

The combination of drifts taken at different epochs with small variations in calibration, the ``blind'' baselining done 
during baseline subtraction and the drift nature of the data acquisition produce various systematic blemishes in the spectral 
grids. Partial correction of those blemishes is achieved by (a) re-baselining the gridded data along the spectral
dimension and (b) performing a similar task in the spatial dimensions, something akin to the flatfielding of optical images. 
For most grids, these procedures were performed using standard automated routines, discussed further in Appendix \ref{app:pipeline}.

\subsection{Source identification}\label{sec:sourceid}

Signal extraction is initiated on the fully processed 3-D spectral grids
in the Fourier domain with an automated matched
filter algorithm described by \citet{saintonge07a} to produce a list of candidate HI detections.
Each grid and its associated candidate catalog are then inspected together in a visualization 
environment called GridView which allows users to manipulate the spectral grids as well as to overlay datasets
from various redshift catalogs and databases such as the Sloan Digital Sky Survey 
\citep[SDSS;][]{york00a}, SkyView \citep{mcglynn94a}, the Second Palomar Observatory Digital
Sky Survey \citep[POSS-II; ][]{djorgovski98a}, and the NASA Extragalactic Database (NED).
In the vicinity of candidate detections, both polarizations, existing catalog entries and redshifts,
and imaging databases are examined closely to identify, where applicable, the most probable
optical counterpart (OC), and possibly to corroborate (or reject) the candidate HI line signal. 
The process of identification of OCs is described in Section 4.1 of \citet{haynes11a}. The close examination
of the ALFALFA grids as well as other relevant databases
allows the identification with a higher than normal degree of confidence of 
low-signal-to-noise (SNR) sources which coincide in both space and redshift with likely optical counterparts.
These latter sources, designated as the ``priors'', are useful because they probe fainter flux
levels, as illustrated in Figure 12 of \citet{haynes11a}.

\subsection{Parameter measurement}\label{sec:galflux}

The ALFALFA source catalogs have been produced in a uniform manner to yield a set
of consistent parameters for each object within and across grids. Once an HI source is identified, 
a thumbnail grid is then extracted from a subgrid covering 
at least  $7$\arcmin~$\times~7$\arcmin~
and imported into a measurement environment called GalFlux.

The algorithm used in the GalFlux program  to measure the HI line flux, systemic velocity and
velocity width is depicted in \citet{springob05b} and described 
by \citet{haynes99a}. For a given integrated spectral profile, peak flux density levels $f_{p}$ of each spectral horn 
are selected by eye.  A zeroth moment
map is created by integrating the flux over these channels with pixel values given by
\begin{equation}
M_{0}=\int S_{\nu} dv~~\rm{[mJy~beam^{-1}~km~s^{-1}]}
\end{equation}
Ellipses are fit to the moment map image at multiple isophotal levels; for the catalogued
measurements presented here, the isophotal half-peak intensity level
for the moment map is used; 
see Section \ref{app:gridapp} for further details.

The total HI line flux density is derived from the source image, taking into account
the telescope beam pattern \citep{shostak80a, kent11a}. The 
source image is spatially integrated over the solid angle covered by the pixels within the 
chosen isophotal fit. This summed flux density $s_{\nu}$ is divided by the sum of the 
beam values $B$ sampled at the position of the image pixels, given by

\begin{equation}\label{eq:beampatt}
S_{\nu}=\frac{\sum_{x_{0}}\sum_{y_{0}}s_{\nu}(\Delta x, \Delta y)}{\sum_{x_{0}}\sum_{y_{0}}B(\Delta x, \Delta y)}~~\rm{[mJy]}
\end{equation}

The HI line flux density S$_{21}$ is then summed over all velocity channels containing signal and a statistical
error is estimated. 
It is important to note that this HI line flux density measurement technique 
is optimized for source sizes of approximately the beam area or smaller. 
\citet{haynes11a} performed a comparison of the HI flux densities measured by ALFALFA versus
those reported by \citet{springob05a} and by HIPASS \citep{meyer04a}. We have repeated the comparisons
for the larger sample reported here with statistically indistinguishable results; analysis of
differences is complicated by the various corrections required to account for pointing and
source extent in the \citet{springob05a} sample and by SNR effects.
While ALFALFA is a blind mapping survey and should recover all of the flux for extended sources,
the pipeline processing employed in producing the catalog presented here may miss flux from
the most extended or highly asymmetric sources. The catalogued
HI line flux densities given in Table \ref{tab:exgalcat} are derived over the area encompassed by
the isophote at 50\% of the peak power. This isophote is typically comparable to or larger than
the beam (at the full width at half maximum power: FWHM) while the vast 
majority of sources are unresolved.
The integrated flux densities for very extended sources or with
significant angular asymmetries can be misestimated by the pipeline 
algorithm.
A special catalog with parameters
of extended sources is in the process of construction (Hoffman {\it et al.} in prep).

The global HI line velocity and velocity width provide measures of the galaxy's
systemic recessional velocity and projected disk rotational velocity. The 
optimal definition of particularly the width measurement
depends in part on the science objective (redshift, measures of rotational
velocity) and data quality factors such as the impact of turbulence 
and SNR \citep[e.g][]{schneider86a, bicay86a, chengalur93a, catinella07a}.
The algorithm used here is nearly identical to that presented by
\citet{chengalur93a} and \citet{springob05b}.
To derive a measure of the profile width, polynomials (usually lines) are fit
to the channels at both edges of the emission between 15\% and 85\% of the
peak flux. The velocity width $W_{50}$ is then defined as the difference between the
velocities corresponding to the fitted polynomial at a level
of 50\% of the maximum value of flux
on each horn. In practice, the maximum flux
value is adopted as the observed peak flux minus the
rms, $f_{p}$-rms, in order to correct for the contribution of noise.   
The average of the two velocities at 50\%  of $f_{p}$-rms is taken as
the systemic velocity $V_{50}$. 
Similar measurements are also made at the 20\% level for the velocity
and associated width.  

In addition to statistical uncertainties in the integrated HI flux density, velocity 
and velocity width, a subjective estimate of an additional systematic uncertainty is
obtained by examination of the minimum and maximum extents of the HI emission signal
(see \ref{app:gridapp}). In most cases, the statistical uncertainty is larger than
the systematic one and the latter can be ignored. However, in cases of low SNR,
very narrow velocity width and/or shallow outer profile slope, the statistical error
is clearly too small and thus the adopted uncertainty is the sum in quadrature of the two.
Velocity width measurements play an
important part in determining the signal--to--noise ratio (SNR) 
of a spectrum, especially for sources with $W_{50}~>$~400~\kms~\citep{haynes11a}.

Velocities and velocity widths, 
centroid sky positions from the isophotal ellipse fits, 
SNR, etc., are all measured and stored in individual
source files with the spectral profiles.  Optical identification is also made
based on previous HI observations \citep{springob05b} and visual
inspection of the imaging databases from the SDSS, POSS-II, and Two Micron All Sky Survey
\citep[2MASS;][]{skrutskie06a}.
Each source file can be reviewed in a catalog viewing program with a complete
history of how the source was measured and the kinematic parameters derived.
This tool, called GalCat (for Galaxy Catalog), connects to sites
that use uniform International Virtual Observatory Alliance (IVOA)
web service standard protocols \citep{graham07a}.

Previous ALFALFA data releases have included a category of objects without OCs which lie within the velocity
regime associated with the Galactic HI emission including the well-known HI high velocity clouds (HVCs).
HVCs are often very extended, exceeding in size that of a single ALFALFA grid. For such objects,
ALFALFA may identify multiple sources associated with a single cloud or complex, and
only the flux in the knots of emission will be measured. A very few of the most compact HVCs dubbed the
``Ultra-Compact'' HVCs (UCHVCs) prove to be very nearby galaxies in or near the Local Group 
\citep{giovanelli10a, giovanelli13a,
adams15a, janesh15a, janesh17a}, but the majority are likely associated with Galactic phenomena. Because
ALFALFA was not optimized to probe the velocity range associated with Galactic HI, the
HVCs identified by the standard ALFALFA pipeline must be interpreted with understanding 
of its limitations \citep{adams13a, bianchi17a}. For this reason, they are not included here.

\section{The ALFALFA Extragalactic Source Catalog}
\label{sec:catalog}

\subsection{Extragalactic HI Sources}

Table \ref{tab:exgalcat} presents the principal results of the ALFALFA extragalactic HI survey. 
The contents of Table \ref{tab:exgalcat} are largely similar to those presented in previous ALFALFA
data release catalogs and are as follows:

\begin{itemize}
\item {Col. 1: Entry number in the Arecibo General Catalog (AGC), a private database
   of extragalactic objects maintained by M.P.H. and R.G. The AGC entry 
   corresponds to both the HI line source and the OC where one is assigned. 
   In the absence of a feasible OC, the AGC number corresponds only to the
   HI detection. An AGC number is assigned to all ALFALFA sources; it is intended to be used as
   the basic cross reference for identifying and tracking ALFALFA sources as new data acquired in overlapping
   regions has superceded older results. Note that in previous ALFALFA catalogs, an index number was used, 
   a practice no longer employed. The designation of an 
   ALFALFA source referring only to its HI emission (without regard to its OC)
   should be given using the prefix ``HI'' followed by
   the position of the HI centroid as given in Col. 3 of Table \ref{tab:exgalcat}.
}

\item {Col. 2: Common name of the associated OC, where applicable. Further discussion of the process of
   assigning optical counterparts has been discussed in Section 4.1 of \citet{haynes11a}.
}

\item {Col. 3: Centroid (J2000) of the HI line source, in hhmmss.sSddmmss, after correction for systematic 
   telescope pointing errors, which are on the order of 20\arcsec \ and depend on declination. The systematic
   pointing corrections are derived from an astrometric solution for the NVSS
   radio continuum sources  \citep{condon98} found in the grids.  As discussed in \citet{giovanelli07a} and 
   \citet{kent08a}, the assessment of centroiding errors is complicated
   by the nature of 3-D grid construction from the 2-D drift scans that were often acquired in widely
   separated observing runs, and, for resolved/confused sources of unknown source structure. 
}
   
\item {Col. 4: Centroid (J2000) of the most probable OC, in hhmmss.sSddmmss, associated with 
        the HI line source, where applicable.  The OC has been identified and its likelihood
        assessed following the process discussed in Section 4.1 of \citet{haynes11a}. The median positional 
        offset of the OC from the HI centroid is about 18\arcsec\ and depends on SNR following 
        Equation 1 of \citet{haynes11a}. In rare low SNR instances, it can exceed 1\arcmin.
        It should be noted that only one OC is assigned per HI source although, in some cases, 
        confusion caused by multiple sources (either HI or OCs) within the telescope beam is a possibility.  
}    

\item {Col. 5: Heliocentric velocity of the HI line source, c$z_{\odot}$ in \kms~ 
in the observed frame, measured as the midpoint 
	between the channels at which the line flux density drops to 50\% of 
	each of the two peaks on the low and high velocity horns of the profile
       (or of one, if only one is present); see also \citet{springob05b}. Values adopt the optical convention
        $\delta\lambda/\lambda$, not the ``radio'' one ($\delta\nu/\nu$).
        The statistical uncertainty on c$z_\odot$
	to be adopted is half the error on the width $W_{50}$ tabulated in Col. 6.
}

\item {Col. 6: Velocity width of the HI line profile, $W_{50}$ in \kms, measured at the 50\%
	level of each of the two peaks, as described in Col. 5 and corrected
	for instrumental broadening 
        following Equation 1 of \citet{springob05a}. No corrections due to
	turbulent motions, disk inclination or cosmological effects are
	applied. The estimated uncertainty on $W_{50}$, $\sigma_W$, in \kms, follows
        in parentheses. This error is the sum in quadrature of two components: a
	statistical error dependent on the SNR of the HI signal, and a systematic error 
        associated with the user's confidence in the definition of spectral boundaries of the signal
        and the applied baseline fit; see \ref{sec:galflux} and \ref{app:gridapp}.
        In the majority of cases,
	the systematic error is smaller than the statistical
	error and can be ignored.
}

\item {Col. 7: Velocity width of the HI line profile, $W_{20}$ in \kms, similar to $W_{50}$, 
        but measured at the 20\%
	level of each of the two peaks. Note that the algorithm used to determine widths by fitting 
        a polynomial to each horn between 15\% and 85\% of the peak flux 
        is optimized to
        measure $W_{50}$ and $\sigma_W$ at 50\% of the peak, not at the lower value of 20\%.
        While some authors prefer to use the lower value, we find that it is less robust
        and its error harder to quantify, particularly at lower SNRs. 
} 

\item {Col. 8: Integrated HI line flux density of the source, $S_{21}$, in Jy~\kms. The estimated uncertainty 
        of the integrated flux density $\sigma_{S_{21}}$, in Jy \kms, 
	is given in parentheses and has been derived following the same procedure as used to measure the
        uncertainty in the velocity width $W_{50}$ (Col. 6). As discussed in Section \ref{sec:galflux} and the 
        Appendix \ref{app:gridapp},
        line flux density values included in Table \ref{tab:exgalcat} have been extracted 
        from the spatial integration of the 3-D grid
        over a window of at least  $7$\arcmin $\times 7$\arcmin and corrected for the survey beam
        over the same area. The algorithm used may underestimate the flux of very extended 
        and/or asymmetric sources.
}

\item {Col. 9: Signal--to--noise ratio SNR of the detection, estimated as 
   \begin{equation}
	SNR=\left (~\frac{1000S_{21}}{W_{50}} \right ) \frac{w_{smo}^{1/2}}{\sigma_{rms}}
	\label{eq:eqsn}
	\end{equation}
   where $S_{21}$ is the integrated flux density in Jy \kms, as listed in Col. 8.
   The ratio $1000 S_{21}/W_{50}$ is the mean flux density across the feature in mJy.
   In this definition of SNR, $w_{smo}$ is a
   smoothing width expressed as the number of spectral resolution
   bins of 10 \kms \ bridging half of the signal width and $\sigma_{rms}$
   is the rms noise figure across the spectrum measured in mJy at 10
	\kms \ resolution, as tabulated in Col. 10. The ALFALFA raw spectra are
   sampled at 24.4 kHz $\sim$ 5.5 \kms ~at $z\sim0$, and, as in previous
   ALFALFA data releases \citep[e.g.][]{giovanelli07a}, $w_{smo}$ is adopted
   as either $W_{50}/(2\times 10)$ for $W_{50}<400$ \kms \ or
   $400/(2\times 10)=20$ for $W_{50} \geq 400$ \kms.
}

\item {Col. 10: Noise figure of the spatially integrated spectral profile, $\sigma_{rms}$,
	in mJy. The noise figure as tabulated is the rms as measured over the signal-- and
	RFI--free portions of the spectrum, after Hanning smoothing to a spectral
	resolution of 10 \kms.
}

\item {Col. 11: Adopted distance $D_{H}$ and its uncertainty $\sigma_D$, both in Mpc. 
        For objects with c$z_{\odot} > 6000$ \kms, 
	the distance is simply estimated as c$z_{cmb}/H_\circ$ where c$z_{cmb}$ is the recessional velocity
	measured in the Cosmic Microwave Background reference frame \citep{Lineweaver96} and $H_\circ$ is
	the Hubble constant, adopted to be 70 \kms~Mpc$^{-1}$. For objects with
	c$z_{cmb} < 6000$ \kms, we use the local peculiar velocity model of \citet{masters05a}, 
        which is based in large part 
        on the SFI++ catalog of galaxies \citep{springob07a} and results
	from analysis of the peculiar motions of galaxies, groups, and clusters, using a combination of primary
	distances from the literature and secondary distances from the Tully-Fisher relation. The resulting model
	includes two attractors, with infall onto the Virgo Cluster and the Hydra-Centaurus Supercluster, as well as
	a quadrupole and a dipole component. The transition from one distance estimation method to the other
	is selected to be at c$z_{\odot}=6000$ \kms \ because the uncertainties in each method become 
        comparable at that distance.
	Where available, primary distances in the published literature are adopted; we also use secondary distances,
        mainly from \citet{tully13a}, for galaxies with c$z_{cmb} <$ 6000 \kms. When the galaxy
	is a known member of a group \citep{springob07a}, the group systemic recessional velocity c$z_{cmb}$ is 
        used to determine the distance
	estimate according to the general prescription just described. Where primary distances are not available,
        objects in the Virgo region are assigned
        to likely Virgo substructures \citep{mei07a} and then the distances to those subclusters are adopted 
        following \citet{hallenbeck12a}. Errors on the distance are generated by running 1000 Monte Carlo iterations 
        where peculiar velocities are drawn from a normal distribution each time as described in Sec. 4.1 
        of \citet{jones18b}. Such errors are likely underestimates in the vicinity
        of major attractors such as Virgo. 
        It should be noted that the values quoted here are Hubble distances, 
        not co-moving or luminosity distances.
}
	
\item {Col. 12: Logarithm of the HI mass $M_{HI}$, in solar units, computed via the standard
	formula $M_{HI}=2.356\times 10^5 D_{H}^2 S_{21}$ and assuming the distance given in Col. 11
        (not the luminosity distance). The uncertainty $\sigma_{\log M_{HI}}$ is
        derived, following \citet{jones18b}, by combining the uncertainty in the integrated HI line flux and 
        the distance with a minimum of 10\% uncertainty. The latter minimum is set to prevent the 
        error from ever getting unrealistically small and to include the systematic uncertainty in the 
        flux calibration. 
        The uncertainty in $\log M_{HI}$ then is
\begin{equation}
        \sigma_{\log M_{HI}} = \frac{\sqrt{\left(\frac{\sigma_{S_{21}}}{S_{21}}\right)^{2} + 
                  \left(\frac{2\sigma_{D}}{D}\right)^{2} + 0.1^{2}}}{\ln{10}}
\end{equation}
        It should be noted that the HI mass values given here do not include a correction for HI self-absorption.      
} 

\item {Col. 13: The HI source detection category code, used to distinguish the high SNR sources from the lower ones
        associated with OCs of comparable redshift. 
	\hskip 8pt Code 1 refers to the 25434
        sources of highest quality. Quality is assessed based on
        several indicators: there is a good match in signal characteristics
        match between the two independent polarizations observed by ALFALFA,
	a spatial extent consistent with the telescope beam (or larger), an RFI-free spectral profile,
	and an approximate minimum SNR threshold of 6.5 \citep{saintonge07a}. These exclusion
	criteria lead to the rejection of some candidate detections with SNR $>6.5$; likewise, some
	features with SNR slightly below this soft threshold are included because of optimal overall characteristics
	of the feature, such as well-defined spatial extent, broad velocity width, and obvious association with an OC. 
        We estimate that the detections with code 1 in Table \ref{tab:exgalcat} and associated with an OC
        are nearly 100\% reliable; the completeness and reliability of
        the $\alpha.40$ catalog are discussed in Section 6 of \citet{haynes11a}.

	\hskip 8pt Code 2 refers to the 6068 sources categorized as ``priors''. They are sources of low 
        SNR ($\lesssim$ 6.5),  
        which would ordinarily not be considered
	reliable detections by the criteria set for code 1, but which have been matched with OCs with
	known optical redshifts coincident (to within their uncertainties) with those measured in the HI line. We include 
        them in our catalog because they are very likely to be real. 
        As defined, the classification of a source as a prior depends not just on the redshift match with a likely counterpart
        but also on the profile shape, polarization match and location relative
        to RFI so that the SNR as defined is not a good statistical indicator of their reliability.
        In fact, 15 of the priors have SNR $<$ 3. In general, the priors should not be used in 
        statistical studies which require well-defined completeness limits; this point is further discussed in
        Section 6 of \citet{haynes11a}). Because of the substantially more complete SDSS spectroscopic coverage of the
        ALFALFA region in the northern Galactic hemisphere than the southern, the number of priors is substantially
        higher in the former than the latter.
        
        It should be noted that objects without optical counterparts and c$z$ in the range of Galactic HI emission, including the high velocity clouds, are not included in Table \ref{tab:exgalcat}; in previous ALFALFA data releases, such sources have been included and identified as HI source category ``9'' sources. They are discussed in \citet{giovanelli10a, adams13a} and a complete catalog will be presented elsewhere.
}

\end{itemize}

\onecolumngrid
\floattable
\begin{deluxetable}{rcccrrrrrrrrc}
\rotate
\tablecolumns{13}
\tablecaption{ALFALFA Extragalactic HI Source Catalog\label{tab:exgalcat}}
\tablehead{
\colhead{AGC ID} & \colhead{Name} & \colhead{HI position} & \colhead{OC position} & \colhead{c$z_{\odot}$} &
\colhead{W$_{50}$~~}  & \colhead{W$_{20}$} & \colhead{$\int SdV$~~} & \colhead{SNR} & 
\colhead{rms} & \colhead{D$_H$} & \colhead{logM$_{\rm HI}$} & \colhead{Code} \\
\colhead{ }     &  \colhead{ }    &  \colhead{J2000} &  \colhead{J2000}  &  \colhead{\kms}  & 
\colhead{\kms}   & \colhead{\kms} &  \colhead{Jy-\kms} & \colhead{ }     &  
\colhead{mJy} &  \colhead{Mpc}  & \colhead{logM$_{\odot}$}  &  \colhead{ }\\
\colhead{(1)} &  \colhead{(2)}  & \colhead{(3)} & \colhead{(4)} & \colhead{(5)} & 
\colhead{(6)} &  \colhead{(7)}  & \colhead{(8)} & \colhead{(9)} & 
\colhead{(10)} &  \colhead{(11)} & \colhead{(12)} & \colhead{(13)}
}
\startdata
105367 &          & 000000.4+052636 & 000000.8+052633 & 11983 & 274(~39) & 281 &    1.14(0.08) &   8.1 &   1.91 & 166.0( 2.3) &  9.87(0.05) & 1\\
333313 &          & 000000.9+245432 & 235959.4+245427 & 11181 & 313(~20) & 333 &    1.80(0.09) &  11.3 &   2.02 & 154.8( 2.3) & 10.01(0.05) & 1\\
331060 & 478-009b & 000002.5+230505 & 000003.4+230515 &  4463 & 160(~~4) & 184 &    1.96(0.07) &  14.7 &   2.35 &  50.6(10.4) &  9.07(0.18) & 1\\
331061 & 456-013  & 000002.5+155220 & 000002.1+155254 &  6007 & 260(~45) & 268 &    1.13(0.09) &   6.5 &   2.40 &  85.2( 2.4) &  9.29(0.06) & 1\\
104570 &          & 000001.6+324230 & 000001.2+324237 & 10614 & 245(~~6) & 250 &    0.86(0.07) &   6.6 &   1.86 & 147.0( 2.3) &  9.64(0.06) & 1\\
331405 &          & 000003.3+260059 & 000003.5+260050 & 10409 & 315(~~8) & 345 &    2.62(0.09) &  16.1 &   2.05 & 143.8( 2.2) & 10.11(0.05) & 1\\
102896 &          & 000006.8+281207 & 000006.0+281207 & 16254 & 406(~17) & 433 &    2.37(0.12) &  11.2 &   2.31 & 227.4( 2.2) & 10.46(0.05) & 1\\
630358 & 382-015  & 000007.5-000249 & 000007.8-000226 &  7089 &  70(~~9) & 103 &    2.47(0.06) &  29.7 &   2.20 &  96.2( 2.3) &  9.73(0.05) & 1\\
105368 &          & 000010.9+041654 & 000011.7+041637 &  3845 &  83(~~6) &  94 &    0.72(0.06) &   7.5 &   2.33 &  54.2( 2.2) &  8.70(0.07) & 1\\
331066 & 382-016  & 000011.5+010723 & 000012.7+010712 &  7370 & 214(~22) & 299 &    2.30(0.11) &  13.3 &   2.64 & 100.2( 2.2) &  9.74(0.05) & 1\\
102571 &          & 000017.2+272359 & 000017.3+272403 &  4654 & 104(~~3) & 124 &    2.00(0.06) &  19.0 &   2.29 &  65.9( 2.1) &  9.31(0.05) & 1\\
102728 &          & 000021.2+310038 & 000021.4+310119 &   566 &  21(~~6) &  36 &    0.31(0.03) &   7.5 &   1.92 &   9.1( 2.2) &  6.78(0.22) & 1\\
331067 & 517-010  & 000020.0+343641 & 000022.2+343658 & 12687 & 104(~13) & 149 &    0.99(0.08) &   7.8 &   2.75 & 176.7( 2.3) &  9.86(0.06) & 1\\
104678 &          & 000022.4+204808 & 000022.3+204748 &  6852 & 190(~~9) & 220 &    2.49(0.11) &  13.6 &   2.95 &  92.9( 2.2) &  9.70(0.05) & 1\\
105370 &          & 000027.7+053256 & 000029.6+053323 & 13133 & 245(~~6) & 252 &    1.04(0.09) &   6.6 &   2.26 & 182.5( 2.3) &  9.91(0.06) & 1\\
 12893 & 456-014  & 000028.0+171315 & 000028.1+171309 &  1105 &  71(~~2) &  85 &    2.30(0.05) &  29.2 &   2.07 &  12.8( 4.4) &  7.95(0.30) & 1\\
 12896 & 478-010  & 000030.1+261928 & 000031.4+261931 &  7653 & 170(~10) & 217 &    3.14(0.08) &  22.0 &   2.44 & 104.5( 2.3) &  9.91(0.05) & 1\\
102729 &          & 000032.1+305152 & 000032.0+305209 &  4618 &  53(~~6) &  71 &    0.70(0.04) &  10.5 &   2.02 &  65.4( 2.1) &  8.85(0.06) & 1\\
331070 &          & 000033.8+224645 & 000033.6+224642 & 11715 &  80(~~3) & 100 &    1.49(0.06) &  16.4 &   2.26 & 162.4( 2.3) &  9.97(0.05) & 1\\
 12895 &          & 000039.6+200333 & 000038.3+200332 &  6746 & 162(~~2) & 181 &    3.82(0.07) &  29.9 &   2.23 &  91.3( 2.3) &  9.88(0.05) & 1\\
\enddata
\begin{flushleft}
Table \ref{tab:exgalcat} is published in its entirety in machine-readable format. A portion is shown here for guidance regarding its form and content. 
\end{flushleft}
\end{deluxetable}

\subsection{HI Spectra of ALFALFA Extragalactic HI Sources}\label{sec:spectra}

In addition to the measured and derived properties presented in Table \ref{tab:exgalcat}, a final representative 2-D HI line spectrum
for each source has been extracted over a window in each grid outlined by the isophote at half of peak intensity. Spectral files are provided both in ASCII and FITS formats and include values for each spectral channel of the frequency, heliocentric velocity (c$z$), flux density, the value of the subtracted polynomial baseline, and the normalized weight. We emphasize that the algorithm used to derive the total HI line flux presented in Table \ref{tab:exgalcat} integrates the source image over the solid angle covered by pixels contained within half-peak isophote and then applies a correction for the beam pattern as indicated in Equation \ref{eq:beampatt}. Caution should be exercised for those channels for which the normalized weight is low, e.g. $<$ 0.5.
These spectra will be available at {\it http://egg.astro.cornell.edu/alfalfa/data/} and through the NASA Extragalactic Database (NED: {\it https://ned.ipac.caltech.edu/}).

\subsection{OHM Megamasers and Candidate OHMs}\label{sec:ohmcand}
As pointed out by \citet{briggs98a},  OH megamasers (OHMs) redshifted from 18 cm rest wavelength into the 
targeted HI bandpass can be a source of contamination in blind HI line surveys of the local Universe. 
For ALFALFA, the relevant OHM redshift range is 0.167 $< z <$ 0.244.
Several OHM candidates were presented in \citet{haynes11a}. In a few cases, the ALFALFA signals have been recognized
as matching the OH emission of previously-known OHMs discovered by \citet{darling06a}. 
\citet{suess16a} conducted a concerted program of optical spectroscopy to asses the likelihood of OHM interlopers.
Here we present in Table \ref{tab:ohmcat}, the small number of HI sources which have been
flagged as possible interloping OH megamasers.
19 sources have been thus identified, nine of which coincide with likely OC's which have known optical redshifts in the 
appropriate redshift; the remaining ten sources have not been confirmed but should be considered as candidate OHMs. All 19
sources are listed in Table \ref{tab:ohmcat}, but separated into confirmed (top) and candidates (bottom). 
Because of the uncertainties, we give only basic parameters for these sources and
reiterate that the final ten OHM candidates need further confirmation. The frequency of the center of the line signal
$f_{sys}$, in MHz, is included along with its approximate velocity width, line flux density, SNR and rms.
Furthermore, 
there may be additional OHMs lurking
but not yet identified in the ALFALFA catalog, consistent with the estimate of 9$_{-6}^{+73}$ found by 
\citet{suess16a}.

\onecolumngrid
\floattable
\begin{deluxetable}{rlcccrrrrl}[ht]
\rotate
\tablecolumns{10}
\tablecaption{ALFALFA OH Megamaser Candidate Catalog\label{tab:ohmcat}}
\tablehead{
\colhead{AGC ID} & \colhead{Name} & \colhead{HI position} & \colhead{OC position} & \colhead{f$_{cent}$} &
\colhead{W$_{50}$~~}  & \colhead{$\int SdV$~~} & \colhead{SNR} & \colhead{rms} & \colhead{$z$} \\
\colhead{ } & \colhead{ } & \colhead{J2000} & \colhead{J2000} & \colhead{MHz} &
\colhead{\kms}  & \colhead{Jy-km/s} & \colhead{ } & \colhead{mJy} & \colhead{ } \\
\colhead{(1)} &  \colhead{(2)}  & \colhead{(3)} & \colhead{(4)} & \colhead{(5)} & 
\colhead{(6)} &  \colhead{(7)}  & \colhead{(8)} & \colhead{(9)} & \colhead{(10)}
}
\startdata
114529 & SDSS J015001.57+240235.8  & 015001.9+240223 & 015001.6+240236 & 1384.12 & 613 & 2.88 &  9.7 & 2.17 & 0.204368 \tablenotemark{a}\\
121379 & IRAS 02524+2046           & 025517.8+205918 & 025517.1+205857 & 1412.20 &  85 & 1.95 & 17.4 & 2.70 & 0.181402 \tablenotemark{b}\\
181310 & SDSS J082312.61+275139.8  & 082311.7+275157 & 082312.7+275138 & 1427.62 &  46 & 2.17 & 15.9 & 2.18 & 0.167830 \tablenotemark{a,b}\\
219215 & SDSS J111125.06+052045.9  & 111126.0+052044 & 111125.1+052046 & 1360.73 &  45 & 0.96 & 16.6 & 1.90 & 0.225213 \tablenotemark{a}\\
219828 & 2MASX J11551476+3130026   & 115518.6+312933 & 115514.7+313003 & 1376.31 & 216 & 0.79 &  4.3 & 2.77 & 0.215989 \tablenotemark{c} \\
229493 & IRAS F12072+3054          & 120949.3+303812 & 120948.3+303750 & 1428.35 & 123 & 1.10 &  9.4 & 2.37 & 0.170000 \tablenotemark{d} \\
229487 & SDSS J120948.28+303749.5  & 121548.9+351149 & 121548.8+351100 & 1415.83 &  35 & 0.49 &  6.5 & 2.80 & 0.166056 \tablenotemark{c} \\
257959 & SDSS J155537.94+143905.6  & 155537.7+143906 & 155537.9+143906 & 1386.22 & 206 & 2.65 & 17.0 & 2.42 & 0.203568 \tablenotemark{a}\\
333320 & IRAS F23129+2548          & 231520.8+260508 & 231521.4+260432 & 1414.25 & 509 & 1.78 &  7.8 & 2.00 & 0.178913 \tablenotemark{b} \\
\hline 
102708 & GALEXASC J000335.98+253204.4 & 000337.0+253215 & 000336.1+253204 & 1426.76 & 234 & 0.91 &  5.7 & 2.33 &  \tablenotemark{e} \\
102850 & 2MASX J00295817+3058322      & 002958.8+305739 & 002958.2+305832 & 1423.24 &  53 & 0.46 &  6.7 & 2.09 &  \tablenotemark{e} \\
114732 & GALEXASC J010107.09+094624.0 & 010110.7+094626 & 010107.1+094621 & 1425.72 &  23 & 0.42 &  7.4 & 2.53 &  \tablenotemark{e} \\
115713 & GALEXASC J014135.31+165731.5 & 014134.1+165718 & 014135.2+165731 & 1424.82 & 133 & 1.36 & 11.6 & 2.27 &  \tablenotemark{e} \\
115018 & GALEXASC J015847.27+073204.2 & 015847.1+073159 & 015847.2+073202 & 1387.23 & 157 & 0.84 &  6.5 & 2.31 &  \tablenotemark{e} \\
124351 & GALEXASC J021750.80+072429.3 & 021751.0+072447 & 021750.9+072428 & 1422.98 &  87 & 0.63 &  5.6 & 2.65 &  \tablenotemark{e} \\
749309 & GALEXASC J101101.08+274012.9 & 101102.9+274020 & 101101.1+274012 & 1400.18 &  80 & 0.72 &  8.1 & 2.20 &  \tablenotemark{e} \\
219835 &                              & 113034.2+322208 & 113034.2+322208 & 1390.52 & 155 & 0.85 &  6.4 & 2.36 &  \tablenotemark{e} \\
249507 & IRAS F14014+3009             & 140341.6+295500 & 140340.3+295456 & 1414.88 & 216 & 1.65 & 12.3 & 2.04 &  \tablenotemark{e} \\
322231 &                              & 223605.9+095743 & 223605.4+095726 & 1429.27 &  40 & 0.55 &  7.9 & 2.41 &  \tablenotemark{e} \\
\enddata
\begin{flushleft}
\tablenotetext{a}{OHM confirmed by \citep{suess16a}.}
\tablenotetext{b}{OHM included in catalog of \citet{darling06a}.}
\tablenotetext{c}{OC has coincident redshift in SDSS.}
\tablenotetext{d}{OHM discovered by \citet{morganti06a}}
\tablenotetext{e}{OC has no reported redshift; OH detection and/or redshift requires confirmation.}
\end{flushleft}
\end{deluxetable}

\section{Summary and Caveats}\label{sec:summ}

This paper presents the catalog of extragalactic HI line sources from the completed ALFALFA HI line
survey. Previous papers, notably \citet{giovanelli05a, saintonge07a, giovanelli07a, kent08a, haynes11a}, have presented
further details on the survey design, observing strategy, signal extraction technique, survey sensitivity and completeness.
Although the minimum-intrusion drift-scan technique attempts to minimize the impact of the complex optics of
the Arecibo telescope and the realities of the terrestrial environment at L-band, the source catalog presented
in Table \ref{tab:exgalcat} should be used with appreciation of numerous caveats. Here we list a few:

\begin{itemize}
\item {HI-selection: The population of galaxies detected by emission of the HI 21cm line is dominated by 
relatively low luminosity, star-forming galaxies.
In fact, virtually all star-forming galaxies contain a cool neutral component of their interstellar medium. 
Therefore the galaxies detected by ALFALFA are preferentially bluer and have lower surface brightness,
lower luminosity and lower metallicity than comparable populations detected by their optical broadband flux.
}

\item {Completeness: Although the ALFALFA source population is 
statistically well behaved as illustrated
by Figure 11 of \citet{haynes11a},
the survey is flux limited in a manner
that depends on the velocity width, e.g. Figure 12 of \citet{haynes11a}. 
Hence completeness corrections need 
to be carefully considered for any statistical analysis for which they are important (e.g., gas
fraction scaling relations).
}

\item {Cosmological corrections: In order to allow immediate comparison with the vast majority of
extant literature and HI line data compilations, we have elected not to apply cosmological corrections to
the values reported in Table \ref{tab:exgalcat}. As a result, velocities are presented in the ``observed''
rest frame, simply as c$z$, the HI line flux densities are given in the commonly adopted hybrid units of Jy-\kms
(as opposed to units of Jy-Hz), and distances are Hubble distances D$_H$. 
Careful discussions of the nature and impact of cosmological corrections
are presented by \citet{hogg99a} and, of particular relevance to HI studies, \citet{meyer17a}.
At large distances, the ``true'' HI mass would be derived use the standard equation \citep[e.g. equation 1 of][]{giovanelli05a}
adopting the luminosity distance D$_L$ and dividing by a factor of $(1+z)^2$ to account for the fact
that the HI flux in observed rest frame units is an overestimate
\citep[e.g. equation 46 of][]{meyer17a}. Because D$_L$ = $(1+z)$~D$_C$ where 
D$_C$ is the co-moving distance, substitution of D$_C$ into the standard equation leads to the factors of $(1+z)$
cancelling out, so that the HI mass can be derived by the standard equation without any $z$ terms but
with the co-moving distance (and the integrated flux in the observed rest frame, as is typical for extant HI surveys) 
instead of the Hubble distance. At the outer edge of ALFALFA, $z$ = 0.06, 
the difference between D$_C$ and D$_H$ is less than 2\% so that the maximum systematic error in the HI mass 
due to using the Hubble distance is $\sim$3\%. Future surveys that explore the Universe beyond
that probed by ALFALFA should be careful to follow the detailed discussion presented in \citet{meyer17a}.
}

\item {Integrated HI line fluxes:
As discussed in Section \ref{sec:galflux} and noted in the description of column 8 of
Table \ref{tab:exgalcat}, the HI line flux densities reported in the catalog here
have been extracted from the spatial integration over a window of at 
least $7$\arcmin $\times 7$\arcmin, with an applied correction factor which models
the beam response pattern
over the same area. 
For this reason, the fluxes derived from the 3-D grids
may underestimate the fluxes of very extended or highly asymmetric sources. Fluxes should match best when the
HI extent is smaller than or comparable to a single ALFA beam.
A special catalog with parameters
of extended sources is in the process of construction (Hoffman {\it et al.} in prep). 
}

\item {Matching with other databases by position: When performing automated matches 
to other catalogs, we strongly advise the use of the OC positions where given. The HI 
centroid positions are 
on average good to only $\sim$20\arcsec ~and their accuracy depends on SNR. 
For low SNR sources, offsets
can exceed 1\arcmin. If the HI position is used in an automated matching, many valid matches 
may be missed or false ones found. At the same time, we admit that some of the OC assignments 
are somewhat subjective, for example, situations where two likely OCs fall within the beam;
some of the assignments made here are certainly incorrect. As further information, particularly
redshifts, becomes available, we invite comment on the current database and 
would plan to provide updated (and improved) versions of the catalog presented here.
}

\item {HI column densities: The majority of sources detected by ALFALFA have
angular sizes much smaller than the telescope beam. While the line flux integral yields
an accurate measure of the HI mass, information on the spatial extent and morphology of
the HI distribution and velocity field is not present. Only in cases of very extended
sources can HI column densities derived from ALFALFA be meaningful. In the absence
of source resolution, measures of the HI column
density reflect the {\it lower limit} on HI column density {\it averaged} over the  
areal extent subtended by the beam. Much higher column density knots could easily be 
present.
}

\item {Assessing the HI extent and distribution: Likewise, except for sources of
angular extent larger than $\sim$5\arcmin, 
follow-up HI synthesis imaging is necessary 
to obtain direct information on the HI column density distribution 
to measure the morphology and inclination of the HI layer, and to derive dynamical
parameters such as the mass contained within the HI radius or the shape of the 
rotation curve. However, it should be noted that synthesis observations can resolve 
out the diffuse, low surface density gas, thus missing flux from extended distributions
that exceed the scale of the shortest interferometer baseline used for such studies.
}

\item {HI self-absorption: Catalogued ALFALFA fluxes include no 
correction for HI self-absorption. Users may wish to implement their own corrections. 
\citet{giovanelli94a} and \citet{jones18b}
both find that the correction for a typical L$^*$ galaxy is likely quite small, 
$\sim$10\%, but might be as high as $\sim$30\% for edge-on galaxies.
}

\item {Global velocity width measures: In Table \ref{tab:exgalcat}, we present HI line widths
measured by fitting a polynomial on both horns of the profile between 15-85\% of the peak
flux on either side. The catalogued values W$_{50}$ and W$_{20}$ give the full widths
at 50\% and 20\% of the peak flux as measured between the polynomial fits
on either side. As has been shown by \citet{bicay86a},
the value of W$_{50}$ is shown to be more robust particularly at lower values of SNR
but other approaches may also be valid. The relationship of the global HI width to
rotational velocity measures obtained from stellar absorption lines or nebular emission
lines is complicated and may depend on galaxy properties such as surface
brightness \citep[e.g.][]{catinella07a}.
}

\item{Corrections to observed widths: The values of W$_{50}$ presented here 
have been corrected for instrumental broadening but not for other factors
such as turbulence \citep{fouque90a} and cosmological stretch.
Furthermore, they reflect only the
projected component of the HI layer's rotational velocity.
}

\item {Distances: While the distance estimation routine represents the best available 
information regarding the distance to each source (that we are aware of), the reality 
is that it employs a highly inhomogeneous collection of primary and secondary distances, 
and group assignments. Furthermore, the flow model itself can be double or triple valued 
in the vicinity of very dense structures. Thus, the distance estimates should 
be considered with caution. \citet{jones18b} find that for calculations with the complete 
catalog the distance estimates are unlikely to result in significant uncertainty or 
bias, but this may not be true for any individual object.
}

\item {HI masses: Under the assumption that the HI is optically thin, we derive
the HI masses simply from the integrated HI line fluxes and the distances. As discussed
in Section \ref{sec:catalog}, the error on log M$_{HI}$ is estimated as the
combination of the uncertainties in the S$_{21}$ and the distance with an additional
allowance for systematic uncertainty in the flux calibration.
As mentioned above, we have applied no correction for HI self-absorption, and
the distances used are Hubble distances. For the most distant ALFALFA
sources, as noted above, the latter effect will introduce a 
bias of similar scale to either the self-absorption or the flux calibration.
The impact of the uncertainties in HI masses is discussed in \citet{jones18b}.
}

\item {Impact of terrestrial interference: RFI impacts some portions of the ALFALFA 
spectrum severely. The spectral data products
for each HI source contain a column with the normalized weight at each individual
frequency/velocity channel; low weight channels should be treated with caution. 
Largely because of contamination by the San Juan airport FAA radar
at 1345-1350 MHz (see Figure \ref{fig:gridrfi}), statistical studies requiring a high degree of volume completeness should be
restricted to galaxies with c$z_{cmb} <$ 15000 \kms.
}

\item {Spectral stacking: The 3-D HI grids used by ALFALFA have demonstrated the power
of spectral stacking \citep{fabello11a, fabello11b, fabello12a, hallenbeck12a, brown15a, odekon16a, brown17a}
to sample selected galaxy populations to low HI mass levels. An important part of the spectral
stacking using ALFALFA has been the retention of the ``weights'' record of missing data and RFI flagging.}

\item {Confusion: As discussed by \citet{giovanelli15a} and \citet{jones16a}, single
dish surveys are most efficient in delivering large statistical samples for
which resolution is not required over volumes where the average separation between
galaxies is larger than the beam area. At even modest distances, the impact of confusion
within the beam needs to be carefully considered. The impact of confusion on the
ALFALFA catalog presented in Table \ref{tab:exgalcat} is estimated to be relatively minor \citep{jones16a},
although individual cases of confused sources are not hard to find.}

\end{itemize}

Statistical studies of the HI-bearing galaxy population still sample relatively small
numbers of galaxies in comparison with spectral surveys at optical wavelengths.
When it was initiated in 2005, ALFALFA followed on the heels of the HI Parkes
Sky Survey \citep[HIPASS;][]{Barnes01a}, the first blind HI survey to cover a large volume.
The much larger collecting area of the 
Arecibo dish offered improvements in sensitivity and angular resolution, advances in
spectrometer capability allowed an increase in spectral bandwidth and resolution, and the 
adopted minimum-intrusion drift scan technique conducted
during nighttime only delivered very high data quality. As a result, the source density of
ALFALFA ($\sim$5 sources per square degree) is more than 25$\times$ higher than that of HIPASS.
The principal aim of ALFALFA has been to survey a wide area of the extragalactic
sky over a cosmologically significant but local volume. Future surveys with single
dish telescopes should focus on deeper surveys of the local Universe or
intensity mapping applications which can actually benefit from lack of
resolution. On-going and planned surveys with 
interferometric arrays will continue to sample the extragalactic HI sky, offering
increased resolution to map the HI distribution and velocity field and extending
beyond the local Universe \citep{giovanelli15a}. 
While this paper presents the ALFALFA harvest, we are confident that the scientific
seeds from ALFALFA promise a future yield that is even more bountiful.

\vskip 20pt

{\bf Acknowledgments}.  The authors acknowledge the work of the entire ALFALFA
collaboration who have contributed to the many aspects of the survey over the years. 
The ALFALFA team at Cornell has been supported by NSF grants AST-0607007,
AST-1107390 and AST-1714828 and by grants from the Brinson Foundation. 
Participation of the Undergraduate ALFALFA Team has been made possible by
NSF grants AST-0724918, AST-0725267, AST-0725380, AST-0902211, AST-0903394, 
AST-1211005, AST-1637339, AST-1637271, AST-1637299, AST-1637262 and AST-1637276. 
EAKA is supported by the WISE research programme, which is financed by the 
Netherlands Organisation for Scientific Research (NWO). 
BRK acknowledges the National Radio Astronomy Observatory (NRAO). The NRAO 
is a facility of the National Science Foundation operated under
cooperative agreement by Associated Universities, Inc. MGJ acknowledges support
from the grant AYA2015-65973-C3-1-R (MINECO/FEDER, UE). We thank Dmitry Makarov 
for comments and suggestions on cross-identifications.

This work is based on observations made with the Arecibo Observatory.
The Arecibo Observatory has been operated by SRI International under a cooperative 
agreement with the National Science Foundation (AST-1100968), and in alliance 
with Ana G. M\`endez-Universidad Metropolitana, and the Universities Space 
Research Association. We thank the staff of the Arecibo Observatory especially 
Phil Perillat, Ganesh Rajagopalan, Arun Venkataraman, Hector H\'ernandez, 
and the telescope operations group for their outstanding support of the 
ALFALFA survey program.

We acknowledge the use of NASA's {\it SkyView} 
facility (http://skyview.gsfc.nasa.gov) located at NASA Goddard Space Flight Center
and the NASA/IPAC Extragalactic Database (NED) which is operated by the Jet 
Propulsion Laboratory, California Institute of Technology, under contract with 
the National Aeronautics and Space Administration. 
The Digitized Sky Surveys were produced at the Space Telescope Science Institute 
under U.S. Government grant NAG W-2166. The images of these surveys are based 
on photographic data obtained using the Oschin Schmidt Telescope on Palomar 
Mountain and the UK Schmidt Telescope. The plates were processed into the 
present compressed digital form with the permission of these institutions. 
The Second Palomar Observatory Sky Survey (POSS-II) was made by the California 
Institute of Technology with funds from the National Science Foundation, the 
National Geographic Society, the Sloan Foundation, the Samuel Oschin Foundation, 
and the Eastman Kodak Corporation. 

This research used data from the Sloan Digital Sky Survey
Funding for the SDSS and SDSS-II has been provided by the Alfred P. Sloan Foundation, 
the Participating Institutions, the National Science Foundation, the U.S. Department 
of Energy, the National Aeronautics and Space Administration, the Japanese 
Monbukagakusho, the Max Planck Society, and the Higher Education Funding Council for England.


\facilities{Arecibo}
\software{IDL}

\bibliography{mybib}

\appendix
\section{Data Pipeline}
\label{app:pipeline}

In this Appendix, we present details of the data reduction pipeline, developed
largely by RG and BK, which has been used to process the ALFALFA survey observational
data and produce its final data products. 

\subsection{Processing of the Drift Scan Data}

The basic data reduction for the ALFALFA survey has
been undertaken in specially-designed software developed in the {\it Interactive Data Language}~(IDL) 
software environment. IDL is
published by Harris Geospatial Solutions and is a dynamically typed, 
single namespace language with advanced graphics
capabilities commonly employed in space science research.  The procedure library
developed by Arecibo staff member Phil Perillat is used in the ALFALFA software reduction pipeline  
for processing data from the WAPP backend.  The pipeline
makes use of the IDL User's Library for data file input/ouput \citep{landsman93a}.

The raw spectral ALFALFA data consist of 1-second records of the individual spectra sampled
by the two separate (linear) polarizations of each of the seven ALFA feed horns via the WAPP
spectrometer.
The initial FITS format files written to disk contain 600 such records, each containing the
spectra for the 2 polarizations of the 7 beams. For processing,
the original FITS files are first converted to
array structures and saved in the native IDL format, so that each ``drift'' 
file is a 2 by 600 by 8 element array.  In each array element, a large
IDL structure is built with the original FITS header information, high-precision
position and time stamp information, as well as a 4096 channel array containing
the raw spectrum.  The 8th array ``beam'' element carries redundant data and is
kept merely for convenience. 

\begin{figure}[t]
\begin{center}
\includegraphics[width=4.0in]{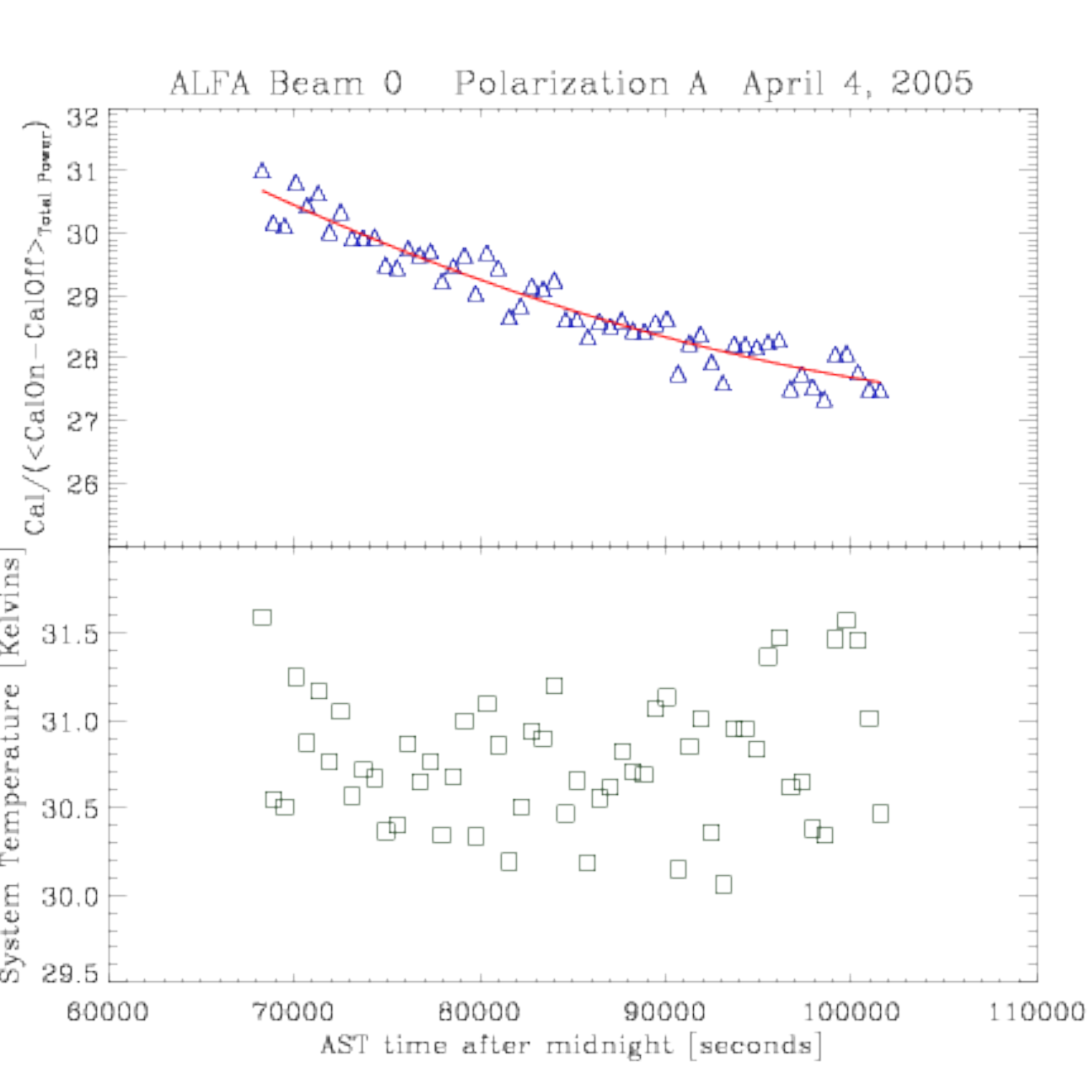}
\caption{Sample calibration solution for one polarization of the center feed horn
for a representative observing session (April 4, 2005).  The upper plot shows the 
calibration ratio Cal/(TP$_{calon}$-TP$_{caloff}$)
and a third order polynomial fit to the data.  The lower plot shows the system temperature
during the observing run. All data shown are for the same polarization (``A'') of the 
central beam (``Beam 0'').}
\label{fig:calib}
\end{center}
\end{figure}

As discussed in \citet{giovanelli05a}, a first stage of calibration is performed by
injecting a noise diode (``cal'') into the system after every 600 seconds of a drift scan.
Each one-second calibration record from
the injected noise diode is written to an identical structure type, 
and saved in its own CALON file.  An ``off'' calibration record is created 
from the final and first records of two adjacent drift files, and
is also saved in its own CALOFF file, yielding a ``triplet'' of associated 
calibration files: the preceding CALOFF, the CALON, and the following CALOFF. 
The series of 600 second drift
and calibration ``triplets'' file lists are created with scan
number prefixes for import into the calibration pipeline.  A temperature calibration ratio
is computed by taking the cal values at 1400 MHz 
as provided and maintained by observatory staff
and dividing by the
difference of the average total power for both the CALON and CALOFF files. 
Corrections for the frequency dependence of the system are applied by the standard
Observatory processing software.
This is performed for each calibration file, beam, and polarization obtained at 600 second
intervals over the whole observing period. Since the telescope is not moved and power levels
are not normally adjusted during an observing run, systematic changes in the calibration ratio
are an indication of gain drift, usually due to electronics. 
A sample calibration ratio for beam 0 and polarization A 
as well as the system temperature $T_{sys}$ during an observing run is shown in Figure~\ref{fig:calib}.
Individual 10 minute calibrations may be affected by weather (lightning), RFI, or the presence
of continuum emission in the CALON/CALOFF scans.
As is evident in that Figure, there is a systematic drift in the calibration ratio, often
observed over the course of the night, particularly in the summer; a common cause is believed
associated with the ability of the dewar to cool the amplifiers as the ambient temperature 
declines during the night. Calibration is applied using the best fit to the observed variation
rather than individual values, thereby reducing the scatter introduced by continuum sources
and RFI in this simplistic approach to calibration.

\begin{figure}[t]
\begin{center}
\includegraphics[width=5.0in]{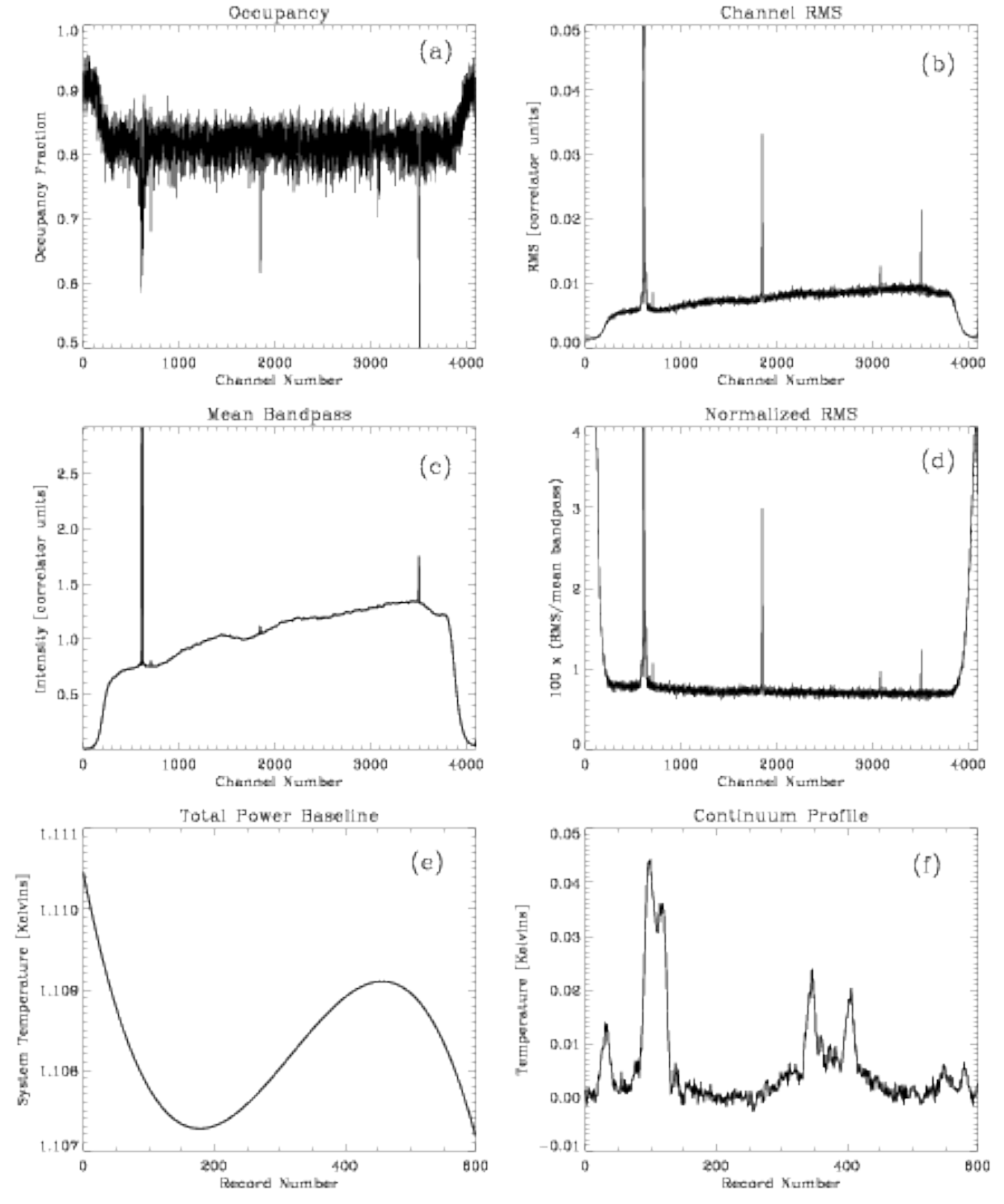}
\caption{Plots showing the products, statistics,
and diagnostics of the bandpass calibration for a single 600 sec drift. 
(a) shows the occupancy fraction
of records that are within 2~$\times$~rms of the time series fit for each channel;
(b) shows the rms for each channel; (c) the mean bandpass of the strip; (d) the rms
divided by the mean bandpass multiplied by a factor of 100; (e) the continuum baseline
contribution after removal of point sources; and (f) the continuum ``strip chart'' integrated
along {\it good} channels.  All data shown are for the same polarization (``A'') of one
of the outer beams (``Beam 4'').}
\label{fig:bpc}
\end{center}
\end{figure}

Bandpass subtraction is performed on 2-dimensional time/frequency plots, one map at a time 
on each polarization and beam for a given drift file.  The calibration process
begins by performing a robust linear fit along the time dimension for each
frequency channel.  The rms is also computed
for each channel, as is the fraction of time series
records less than 2~$\times$~rms.  After exclusion of outliers that deviate more than 2~$\times$~rms 
from the fit, a bandpass value of either (1) the zeroth order coefficient
to the linear fit $c_{0}$, or (2) the median value of the strip is selected.  This option
is chosen by the user at the time of reduction.  Based on experience, option 1 is usually preferable.
The rms as a function of channel is iteratively fit with a 3rd order polynomial.
Channels (including those at the bandpass edges) are flagged that deviate
several standard deviations from the fit.  In addition, channels are
manually flagged, such as those around Galactic HI.  
A cubic spline is used to interpolate the bandpass across the flagged channels .
An ``off'' bandpass is created as the normalized bandpass times the system temperature.
A background total power continuum value is also computed for all time series records for all
records and channels that have not been flagged, excluding point sources; the continuum
contribution from these point sources is also stored away.  Figure~\ref{fig:bpc}~shows the 
multiple diagnostics
of the described calibration process.

\begin{figure}[t]
\begin{center}
\includegraphics[width=3.5in]{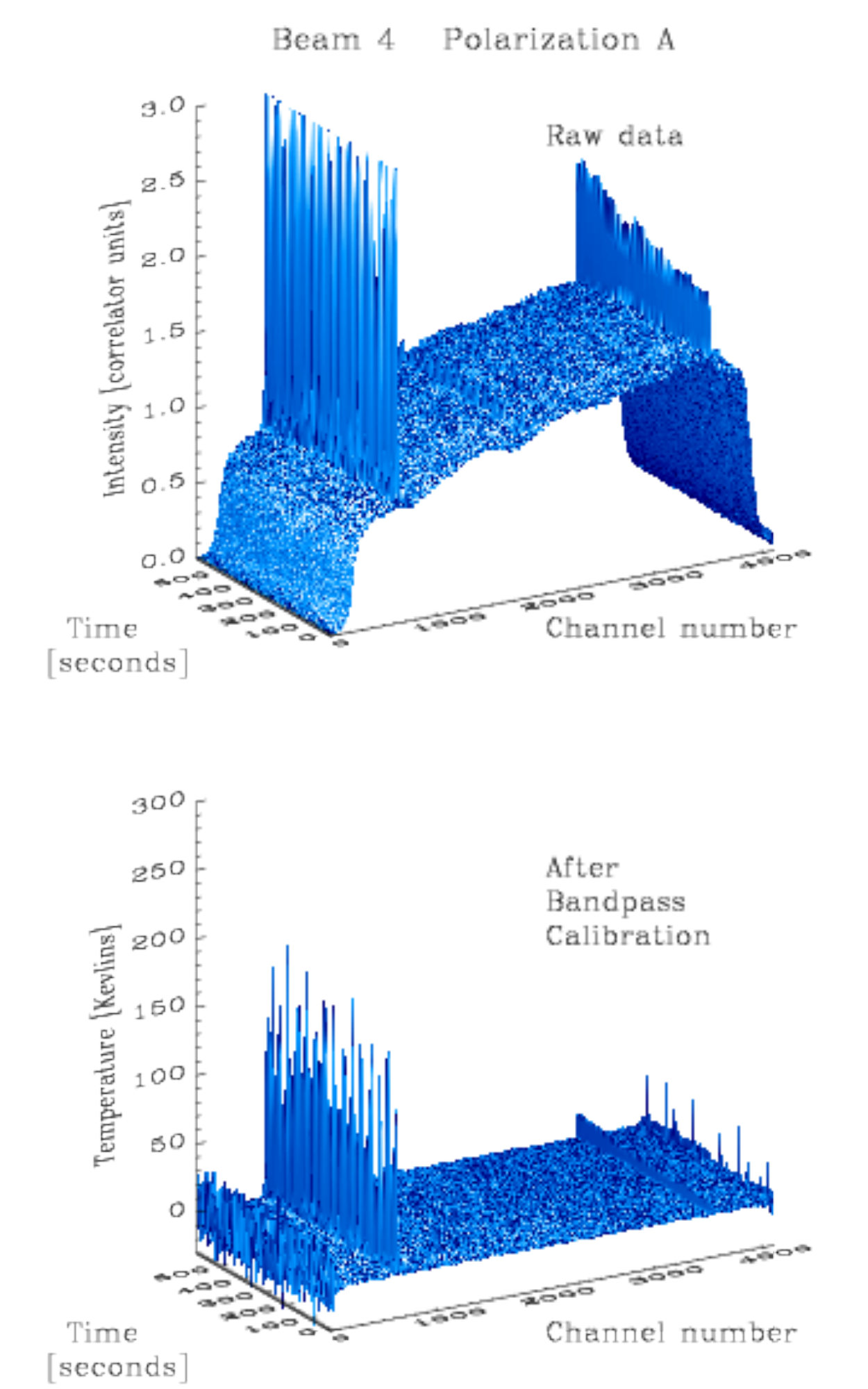}
\caption{2D surface illustrating the bandpass calibration and subtraction process for a 
representative time/frequency (spectrometer ``channel number'') plot for one polarization (''A'') of a single
beam (``Beam 4'') over a 600 sec drift.  The upper plot
shows the raw ALFALFA drift scan spectral data, clipped to an intensity of 3.0 for
dynamic range convenience.  The bottom plot shows the spectral data, scaled to Kelvins,
after the bandpass calibration process.  Features in both plots
include the FAA radar signature near channel 615 (1350 MHz) and the Galactic HI line emission
near channel 3500 (1420 MHz).
\label{fig:levelIplot}}
\end{center}
\end{figure}

The final calibrated and corrected bandpass is computed as 
\begin{equation}
BP_{corr}=\frac{BP_{on}-BP_{off}}{BP_{off}}T_{sys}
\end{equation}
All calibrated values are stored in units of Kelvins.  Calibrated and reduced drift files are saved to disk
for interference flagging.  An example of the process of the bandpass calibration and subtraction process
for a typical time-frequency drift dataset is shown in Figure~\ref{fig:levelIplot}.

\begin{figure}[t]
\begin{center}
\includegraphics[width=6.5in]{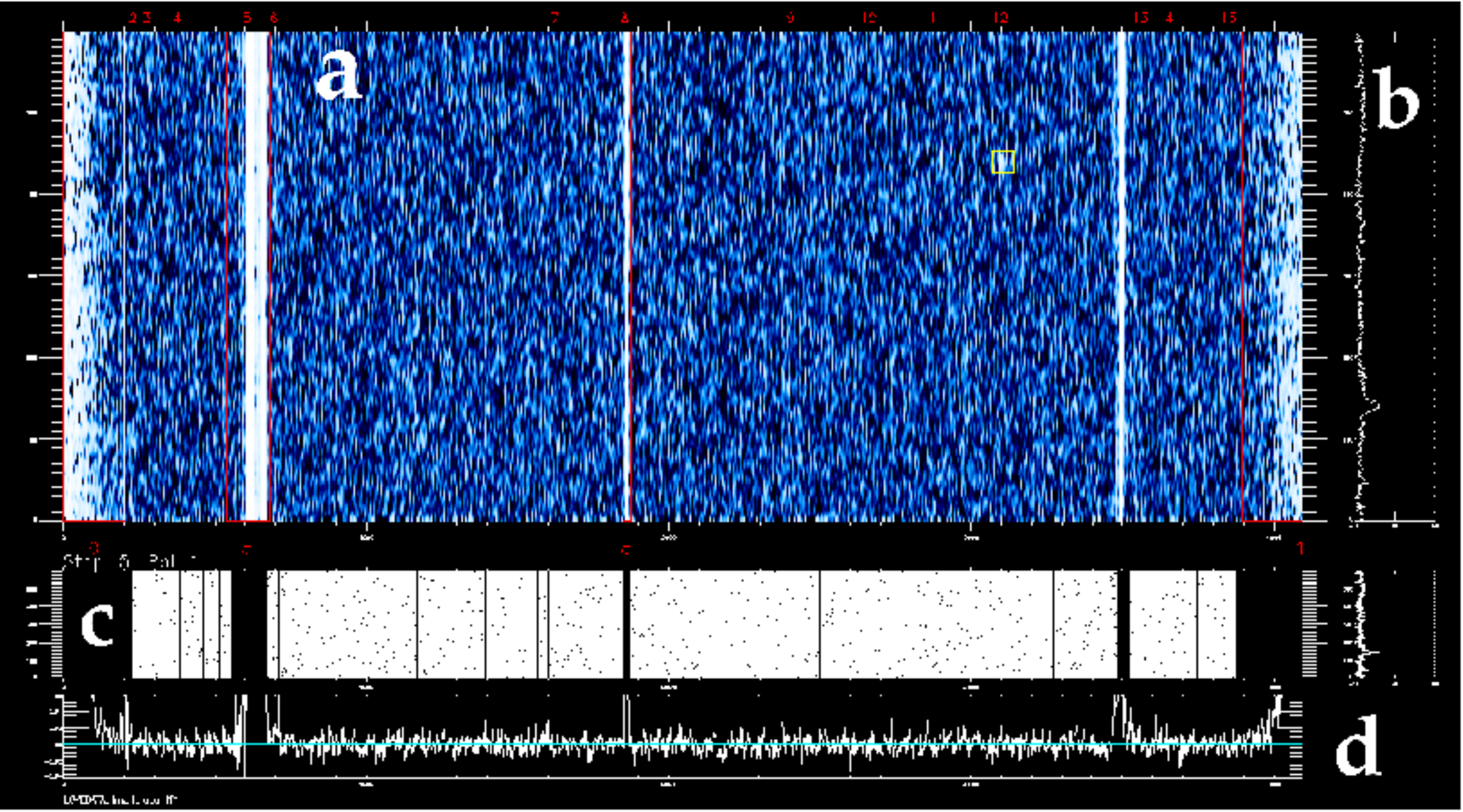}
\caption{Example session of radio-interference flagging from April 4, 2005.  The user
selects regions of interference via predetermined box numbers or user-defined regions.
Cross-referencing with optical and HI databases allows the user to check known information
during the data reduction session.  The figure sections
are as follows: a) the main time vs. frequency plot of the data showing the FAA radar and Galactic HI,
b) the continuum flux integrated along the strip, c) the map mask showing the pixels (in white)
used by the bandpass calibration procedure to produce the continuum flux strip, and d) the average
spectrum for all 600 records.  
Prominent features (indicated by white vertical stripes) include, from left to right,
the strong FAA radar at 1350 MHz, an internally-generated modulation of the radar at 1380 MHz, and
the Galactic HI emission at 1420 MHz.
Red boxes indicated marked regions of RFI, and the yellow box
indicates a galaxy previously identified in the AGC database at the corresponding position and redshift. 
\label{fig:flagbb}}
\end{center}
\end{figure}

Careful attention is given to flagging radio frequency interference for both improvement of data quality
products and to decrease the likelihood of including spurious detections in the automated signal 
extraction process later on.  The RFI flagging stage
is where data quality is first assessed.  Each bandpass calibrated 2D time/frequency plot is 
examined closely by eye.  Areas of interference due to nearby airport radar and associated
harmonics are individually flagged; these records and/or channels are excluded during the signal
extraction and gridding process.  An example flagging session is shown in Figure~\ref{fig:flagbb}.
The data products as this stage in the pipeline are known as Level I data products.	

\subsection{Construction and Analysis of 3-D Grids}\label{app:gridapp}

ALFALFA ``grids'' are 3-D position-position-velocity  ``cubes''
from which final source measurements are obtained.  As discussed in Section
\ref{sec:gridproduction}, the grid centers are located at 
pre-detetermined locations, separated by 8$^{min}$ of R.A.
and centered at Declinations spaced by 2$^\circ$ between +01$^\circ$ and +35$^\circ$.
The process of creating grids begins
by scanning coordinate metadata saved in small size ($\sim$30MB) files containing the
sky positions of every spectrum recorded during each observing
session. A listing is compiled of any 600-second drift files that will contribute to any
position within the specified grid boundaries.
Each drift file is opened and data contributing to the grid are summed in the appropriate R.A./Dec./frequency
bins weighted by a Gaussian kernel of size 2\arcmin.
Grid pixels are approximately 1\arcmin~ square depending on the declination range.  Data are also included
or ignored based on RFI flagging from the level I production process.  A spectral weight map
for each channel is also created based on contributing drifts at each grid point.
The entire process is repeated for total power continuum maps and weights.
Final grids are scaled to units of Jansky beam$^{-1}$ and divided by the appropriate weights
for both spectral and continuum maps.  In addition, because the synthesized beam area
used to generate the grids is larger than that of the telescope at L-band
(from 3.5\arcmin~ to 4\arcmin),
the grids are also multiplied by a gain dilution factor given by
\begin{equation}
G_{dilute}=1+\frac{W_{FWHM}^{~2}}{3^{\prime}.3\times 3^{\prime}.8}
\end{equation}
\noindent where $W_{FWHM}$ is the Gaussian kernel size of 2\arcmin.

Velocity channels are first shifted to the heliocentric velocity frame such that channel
2047 (counting from zero) of all raw spectra are at a frequency of 1385 MHz and c$z_{\odot}$~=~7663~\kms.
The process creates four spectral grids.
Each contains a spectral HI grid of size 144~$\times$~144 pixels
in R.A. and declination, and 1024 channels in frequency space ($\sim$5100 \kms~in c$z$ space).
The four spectral grids are identified by a letter designation
for the c$z_{\odot}$ range that they cover: (a) -2000 to 3300 \kms, (b) 2500 to 7950 \kms, 
(c) 7200 to 12800 \kms,
and (d) 12000 to 17900 \kms.  Each grid overlaps the next by $\sim$1000~\kms~in c$z$ space.  Each of
the four grids occupies approximately 330 MB of disk space, with ancillary files attached containing
a complete history of how the grid was constructed.

Baselining involves fitting polynomials to grid slices, i.e., R.A. vs. spectral channel maps separately 
for both polarizations.  For most maps,
a linear fit is subtracted in the spectral direction.  Special cases
may require excluding Galactic HI, high-velocity clouds, and high SNR extended
detections from the fit.  In areas where residual stray RFI may be present, baselining
is also performed in the R.A. direction.  A secondary process involves subtracting low order
fits to R.A./Dec. maps for each spectral channel.  The subtraction along
each R.A. strip effectively ``flatfields'' the image in channels devoid of any signals.
Extended signals are excluded from the fit, especially for channels containing Galactic HI or 
associated with bright galaxy HI line emission.

\begin{figure}[t]
\begin{center}
\includegraphics[width=6.5in]{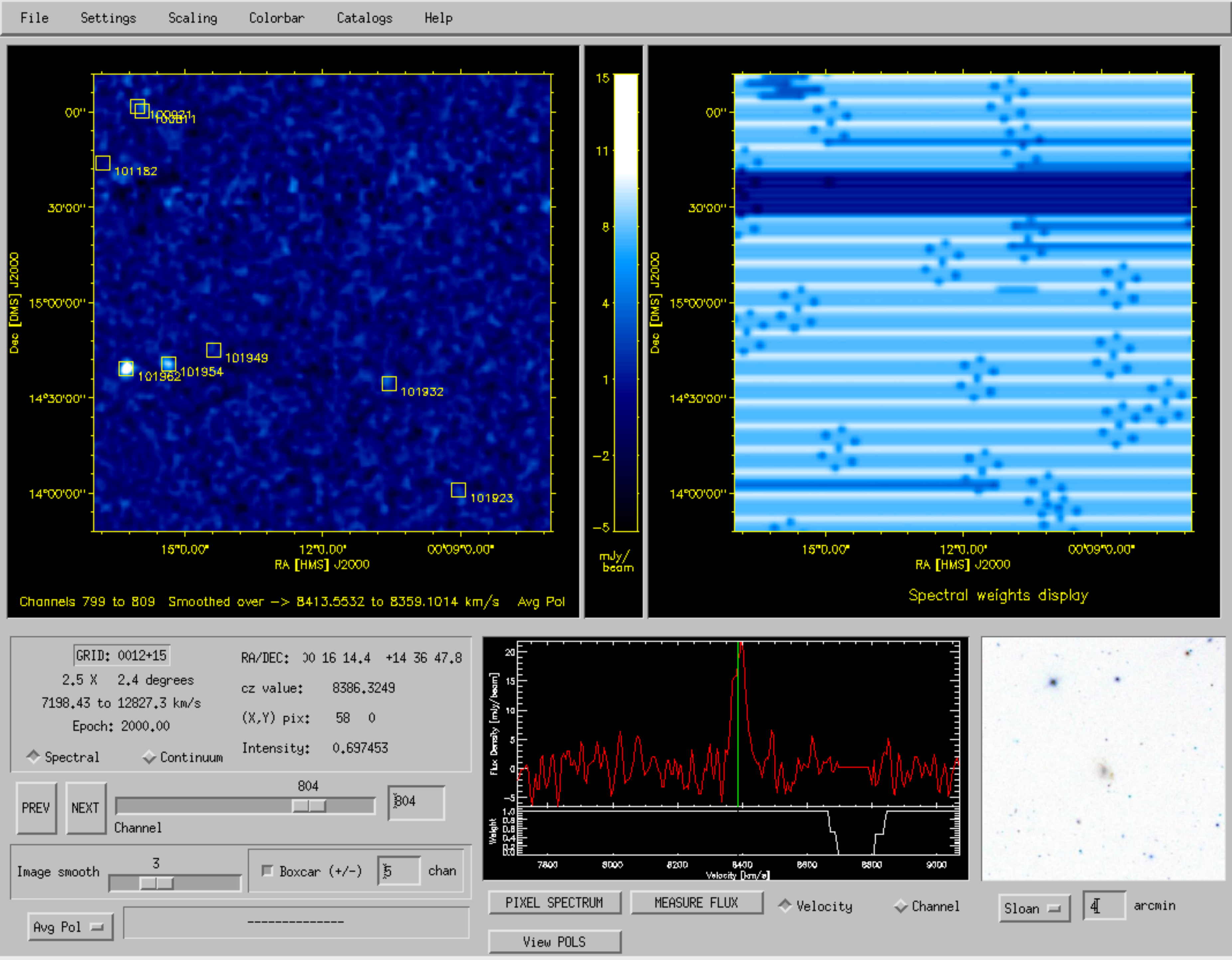}
\caption{GridView application GUI created in IDL. The data cube visualization procedure
incorporates various IVOA web service tools for use during the data analysis process.
The figure shows both optical image viewing and catalog overlay on the
HI line 0th moment map over an adjustable range of channels (left). The display to the
right of the moment map shows the corresponding weight map of the same region; 
horizontal stripes
in the weights map reflect the interleaved drift scan strategy and scalloped nature of 
coverage by beams of different gain. In that display, dark means low weight; there is
a drift missing from this dataset. The repeated pattern of seven darker spots represent the
interruption of data taking for firing of the noise diode (every 600 seconds); the seven spots
are the seven ALFA beams. 
The target under analysis here is AGC 101962, matching
the brightest white spot on the left just below the center of the moment map.
Its spectrum is shown near the center bottom; the spectral weight display, plotted
underneath the flux density indicates that which the spectral range near the
galaxy's HI emission is fine, but that the spectral region around 8700 km/s is of
low weight, due to excision of RFI generated by the NUDET instrument on the GPS constellation
of satellites and an internal modulation of the FAA radar. 
The right bottom window displays the SDSS r-band image of the field
centered on the HI centroid. Users are able to manipulate the 3-D spectral grid
with the various GUI controls, export maps, view polarizations separately, and examine 
the spectrum corresponding to any grid pixel.
\label{fig:gridview}}
\end{center}
\end{figure}

Retained along with the 3-D spectral grid is a continuum map over the same spatial area and a
structure containing the normalized weight of each spectral grid point. The latter provides a 
record of data quality including the ``gain scalloping'' due to differences in the gains
of the central versus peripheral beams, the possibility of missing beams/polarizations
due to hardware failure, RFI excision, and the varying number of drifts covering each spatial point.
Slices in the spectral dimension  (``channel maps'') can be examined along with the 
continuum and comparable ``weights maps'' simultaneous (see Figure \ref{fig:gridview}).
The comparison of grids constructed from separate polarizations provides
a further check on RFI contamination. The retention of such weight information also provides
input for stacking software allowing the identification (and rejection) of sources for
which the data quality do not meet a threshhold for inclusion (e,g, high RFI excision).
The extracted spectra also maintain a value of the normalize weight per spectral channel
for similar reasons.

As mentioned in Section \ref{sec:sourceid}, HI line candidates are identified in
the final grids by applying a matched filter algorithm in the Fourier domain
\citep{saintonge07a}. The candidates are then examined and parameters measured
individually using the tool GridView. An example of a GridView application is given
in Figure \ref{fig:gridview}. Importantly, GridView allows the manipulation (smoothing,
comparing polarizations etc) of channel maps as well as examination of the
weight and continuum maps and the overlay and cross reference of complementary
datasets, redshift catalogs and imaging databases. This functionality is important
for the final catalog data quality and critical for the identification of the
probable OCs. 

Identified sources are then run through the routine GalFlux which measures fluxes and
spectral parameters, allows cross-referencing with other databases and produces
the final data products for each detection. The user selects the region over 
which the spectrum will be
measured. A ``postage stamp'' is extracted, isophotes are fit in the spatial
domain and the spectrum displayed. 
Elliptical isophotes are automatically computed at 
levels of half and one-quarter of the peak power and at levels of 100, 200, 300, 500 and 
1000 mJy beam$^{-1}$; custom isophotes could also be fit. The interactive examination of the 
isophotes allows the user to refine the boundaries in both spatial and spectral 
dimensions and to refine the baseline fit.
The user then can make adjustments to the spatial
and spectral boundaries and can perform additional baseline fitting, smoothing etc. to
determine the best spectral and spatial definition of the HI line signal. Most often,
the metric used to judge the best fit is the SNR; if the
boundaries are too large, noise will drive the SNR down.
In the majority of sources, such choices were 
not important, but this flexibility allows the user to make decisions about the best 
fit in the presence of low SNR, confusion, RFI contamination, polarization mismatch 
or proximity to Galactic HI line emission and to note evidence 
of extended sources or other peculiarity. Such examination
can even result in source rejection; for example, if the feature is very narrow in velocity 
and its spatial definition is inconsistent with the beam pattern, the source is
highly likely to be spurious. The half peak power isophote is typically larger than the ALFA
beam FWHM and should contain all of the flux for sources of extent comparable to or smaller than
the beam. It should be noted, however, that in the vast majority of sources, the true HI
extent is not known but is expected to be smaller than the beam. As noted previously, flux
may be missing from very extended or asymmetric sources when this pipeline approach is
used. A separate work will present fluxes for the known extended sources (Hoffman {\it et al.},
in preparation).

Once the user is confident of the source definition,
then the flux and velocity measurement algorithm, as described in Section \ref{sec:galflux},
is applied to produce a display of the postage stamp, isophotal fits, 
the baselined spectrum and the associated weights over the isophote at 50\%
of the peak flux, and measures of the integrated flux, velocities, velocity widths
and their error. 
In addition to statistical errors on the HI line flux density, velocity and
velocity width, systematic uncertainties on each value are estimated
by flagging minimum and maximum estimates of the extent of the spectral feature and then using
the ratio of those values to estimate uncertainties. In most cases, the systematic uncertainty
is less than the statistical one and can be ignored. In cases of e.g., low SNR, shallow profile 
outer slope or RFI contamination, the statistical error can be clearly too small and the uncertainty
is set as the sum in quadrature of the two.
Optical databases (SDSS, SkyView and NED) can be accessed and displayed
so that the user can assign the most probable OC. The user assigns the detection
category and can enter comments about
the source, the extracted source, the optical environment, etc. The final step
in the analysis of a source is the production of a final output file containing
all of the above information. 

\begin{figure}[ht!]
\begin{center}
\includegraphics[width=6.5in]{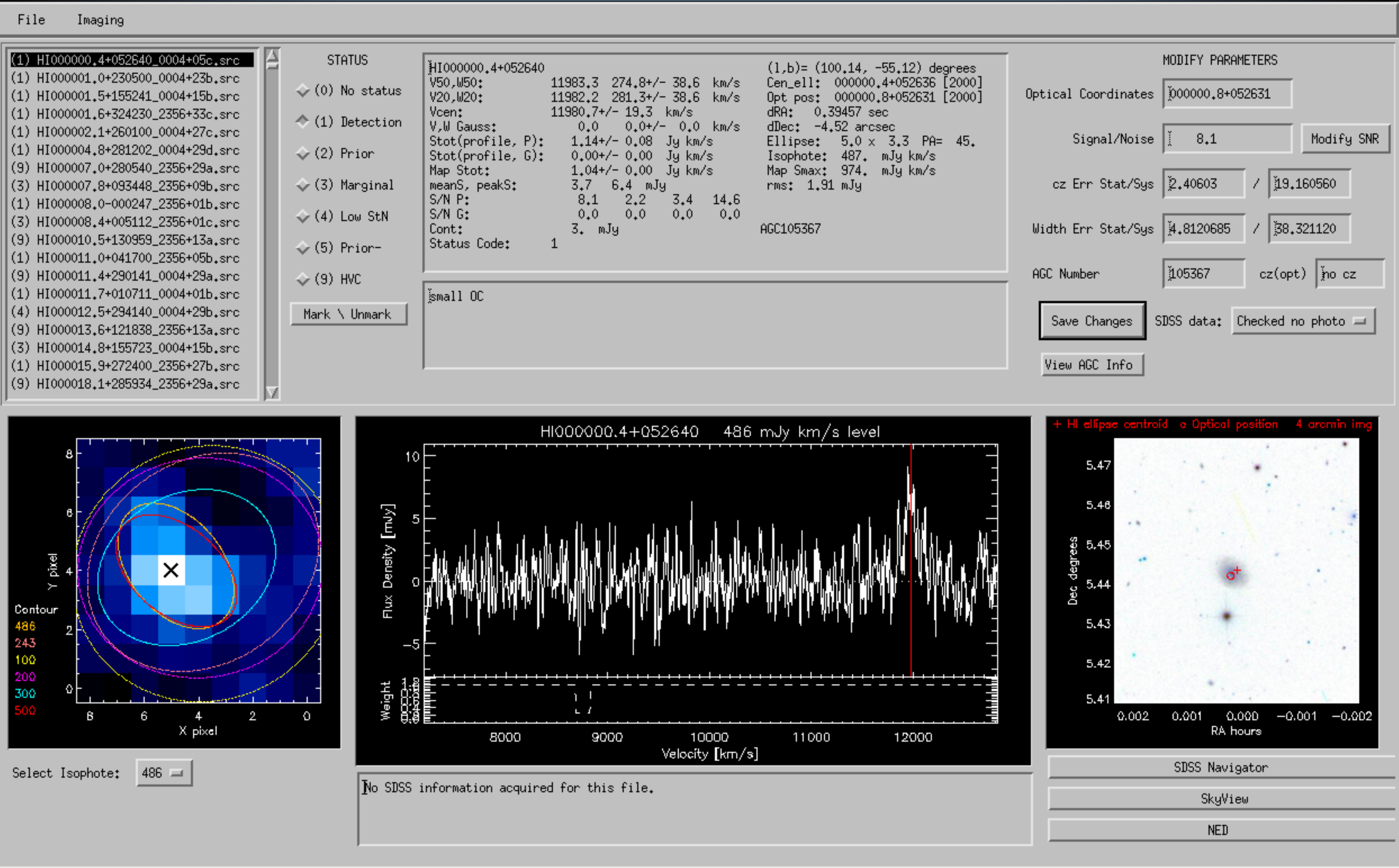}
\caption{GalCat application GUI created in IDL. The catalog of ALFALFA sources is displayed
as a list on the upper left. For each source selected from that list, a summary is 
presented in the interface, including the measured parameters (top center), the
isophotal fits (bottom left), the extracted ``postage stamp'' spectrum and its normalized
weight at each spectral point (bottom center), and the optical image (bottom right). 
The upper right panel offers the possibility to modify parameters as necessary.
\label{fig:srcview}}
\end{center}
\end{figure}

An additional GUI, depicted in Figure \ref{fig:srcview} enables the examination 
of a catalog of HI line sources and their corresponding ``source'' file. Individual
sources can be selected from the catalog listing (upper left); the isophotal fits (lower left),
spectrum and associated weights  (lower center) and corresponding optical field (lower right)
are displayed along with a summary of the derived and assigned parameters. 
A user can update parameters if additional information becomes available; the latter
capability has been an important component over the years.

Additional routines allow production of the catalogs, application of the adopted
flow model \citep{masters05a} and derivation of the HI masses. 
The full ALFALFA data reduction package, dubbed ``Lovedata'' (because we love our data), has
been exported to over 50 sites.

\end{document}